\documentclass[aps,prb,twocolumn,superscriptaddress]{revtex4-1}
\usepackage{amssymb, amsmath}
\usepackage{graphicx,onlyamsmath}

\usepackage{natbib}
\usepackage{xcolor}
\usepackage[normalem]{ulem}

\begin{document}


\title{Hybridization of topological surface states with a flat band}

\author{Sergey S.~Krishtopenko}
\affiliation{Laboratoire Charles Coulomb (L2C), UMR 5221 CNRS-Universit\'{e} de Montpellier, F- 34095 Montpellier, France}

\author{Mauro Antezza}
\affiliation{Laboratoire Charles Coulomb (L2C), UMR 5221 CNRS-Universit\'{e} de Montpellier, F- 34095 Montpellier, France}
\affiliation{Institut Universitaire de France, 1 rue Descartes, F-75231 Paris Cedex 05, France}

\author{Fr\'{e}d\'{e}ric Teppe}
\email[]{frederic.teppe@umontpellier.fr}
\affiliation{Laboratoire Charles Coulomb (L2C), UMR 5221 CNRS-Universit\'{e} de Montpellier, F- 34095 Montpellier, France}
\date{\today}

\begin{abstract}
We address the problem of hybridization between topological surface states and a non-topological flat bulk band. Our model, being a mixture of three-dimensional Bernevig-Hughes-Zhang and two-dimensional pseudospin-1 Hamiltonian, allows explicit treatment of the topological surface state evolution by continuously changing the hybridization between the inverted bands and an additional "parasitic" flat band in the bulk. We show that the hybridization with a flat band lying below the edge of conduction band converts the initial Dirac-like surface states into a branch below and one above the flat band. Our results univocally demonstrate that the upper branch of the topological surface states is formed by Dyakonov-Khaetskii surface states known for HgTe since the 1980s. Additionally we explore an evolution of the surface states and the arising of Fermi arcs in Dirac semimetals when the flat band crosses the conduction band.
\end{abstract}

\maketitle

\section{\label{sec:I}Introduction}
The research on topological materials constitutes one of the most active areas in modern condensed matter physics. Initiated by Kane and Mele, who introduced topology as a new property of two-dimensional (2D) insulators~\cite{w1}, their idea has subsequently been generalized to three-dimensional (3D) materials~\cite{w3}, yielding a whole family of novel topological insulators (TIs). In general, the nontrivial topology of TIs arises from the inversion between two bands with opposite parity, resulting in the appearance of gapless states at the boundaries, which are insensitive to impurities and disorder~\cite{w4,w5,w6,w7}. In 2D TIs counter-propagating one-dimensional (1D) gapless states with opposite spin arise at the edges while 3D TIs feature helical surface states consisting of a single Dirac cone, where the spin points perpendicular to the momentum.

To date multiple 2D and 3D TIs have been experimentally verified (2D TIs~\cite{w8,w9,w10,w12,w13,w14,w26}; 3D TIs~\cite{w15,w16,w17,w18,w19,w20,w21,w22}), yet many materials with inverted band structure are bulk \textit{metals} with additional helical surface states. The most prominent example for this case is the HgTe-class. Here the presence of the heavy-hole $|\Gamma_8,\pm3/2\rangle$ band in combination with an inverted pair of the electron $|\Gamma_6,\pm1/2\rangle$ and light-hole $|\Gamma_8,\pm1/2\rangle$ bands transform HgTe into a bulk semimetal. In order to observe the topologically nontrivial surface states, as predicted by the theoretical analysis based on the $Z_2$ topological invariant~\cite{w23}, it is necessary to open a gap between conduction $|\Gamma_8,\pm1/2\rangle$ and valence $|\Gamma_8,\pm3/2\rangle$ band. In the 3D case this can be achieved by applying tensile biaxial strain to the HgTe bulk film~\cite{w23,w23a} which has been demonstrated experimentally in Ref.~\cite{w24}.

Theoretically the appearance of surface states in 2D and 3D TIs can be understood within the Bernevig-Hughes-Zhang (BHZ) model describing the band inversion in 2D~\cite{w8} and 3D systems~\cite{w16}. However, in addition to the Dirac-like surface states arising from the band inversion, gapless HgTe should also host parabolic \emph{Dyakonov-Khaetskii} (DK) surface states, theoretically predicted in the 1980s~\cite{wDK1}. As the DK states are caused by the coupling between the light-hole $|\Gamma_8,\pm1/2\rangle$ and heavy-hole $|\Gamma_8,\pm3/2\rangle$ band, they should remain even in the presence of the strain-induced band gap~\cite{wDK1a}. Thus, the complete picture of the surface states in strained HgTe should differ significantly from the predictions based on the 3D BHZ model~\cite{w16}. The picture of the surface states becomes even more complex in the case of compressive biaxial strain, when the $|\Gamma_8,\pm1/2\rangle$ and $|\Gamma_8,\pm3/2\rangle$ bands touch at certain points of the Brillouin zone~\cite{w35a}, which is particularly similar to unstrained Cd$_3$As$_2$ crystals~\cite{w32o,w33,w34} known to be 3D Dirac semimetals.

In order to gain better insight into this particular query we investigate \emph{analytically} the transformation of Dirac-like surface states induced by the hybridization between the inverted $|\Gamma_8,\pm1/2\rangle$ and $|\Gamma_6,\pm1/2\rangle$ bands  with an additional heavy-hole $|\Gamma_8,\pm3/2\rangle$ band. To include an additional band, we have combined the 3D BHZ Hamiltonian~\cite{w16} with the 2D pseudospin-1 Dirac-Weyl Hamiltonian (cf. Refs~\cite{w28,w29}) by introducing an effective hybridization strength. This allows us to explore the evolution of topological surface states at different position of the "parasitic" band by varying the hybridization strength. At a specific value of hybridization strength, our linear model qualitatively represents the picture of the surface states in the vicinity of the $\Gamma$ point known from the tight-binding calculations for Cd$_3$As$_2$~\cite{w32} and HgTe~\cite{32HgTe}. The results demonstrate that the parabolic \emph{Dyakonov-Khaetskii} surface states known from HgTe~\cite{wDK1} stem from the modification of the Dirac-like surface states by a hybridization with an additional $|\Gamma_8,\pm3/2\rangle$ band in the bulk.

The paper is structured as follows. We introduce a general analytical model based on a combination of the 3D BHZ and 2D pseudospin-1 Dirac-Weyl Hamiltonian in Section~\ref{sec:II}. Subsequently we discuss the topological surface states at different strengths of the hybridization with the flat band as well as for different boundaries at different position of the flat band, including the 3D TI and Dirac semimetal cases. Finally, the main results are summarized in Section~\ref{sec:IV}.

\section{\label{sec:II}Theoretical model}
For the analytical model we first consider a "modified" 6-band Hamiltonian including a variable hybridization with a "parasitic" bulk band:
\begin{equation}
\label{eq:1}
\hat{H}=\begin{pmatrix}
\hat{H}_{0}(\hat{k}_x,\hat{k}_y,\hat{k}_z) & \hat{H}_{z}(\hat{k}_z) \\ \hat{H}_{z}(\hat{k}_z)^{\dag} & \hat{H}_{0}^{*}(-\hat{k}_x,-\hat{k}_y,-\hat{k}_z)\end{pmatrix},
\end{equation}
Here the asterisk stands for complex conjugation and "$\dag$" corresponds to Hermitian conjugation. The elements $\hat{H}_{0}(\hat{k}_x,\hat{k}_y,\hat{k}_z)$ and $\hat{H}_{z}(\hat{k}_z)$ in Eq.~(\ref{eq:1}) are written as
\begin{equation}
\label{eq:2}
\hat{H}_{0}=\begin{pmatrix}
C_{0}+M_{0} & \hbar v_{\|}\hat{k}_{+}\sin\alpha & \hbar v_{\|}\hat{k}_{-}\cos\alpha \\
\hbar v_{\|}\hat{k}_{-}\sin\alpha & C_{0}+S_{0} & 0\\
\hbar v_{\|}\hat{k}_{+}\cos\alpha & 0 & C_{0}-M_{0}\end{pmatrix}+\mathcal{O}(\textbf{k}^2)
\end{equation}
and
\begin{equation}
\label{eq:3}
\hat{H}_{z}=\begin{pmatrix}
0 & 0 & \hbar v_{\perp}\hat{k}_{z} \\
0 & 0 & 0\\
\hbar v_{\perp}\hat{k}_{z} & 0 & 0\end{pmatrix}+\mathcal{O}(\textbf{k}^2),
\end{equation}
where $\hat{k}_{\pm}=\hat{k}_x\pm i\hat{k}_y$ with $\hat{k}_x$, $\hat{k}_y$, $\hat{k}_z$ being momentum operators. We note that $C_0$ corresponds to a set of zero energies, $S_0$ describes the position of the "parasitic" band and $v_{\|}$ as well as $v_{\perp}$ are the values of velocity for the massless particles. In our case, the $x$, $y$ and $z$ axes are oriented along the (100), (010) and (001) crystallographic directions. For simplicity, we further assume $v_{\|}=v_{\perp}=v$, which can be found in HgCdTe crystals~\cite{w30,w31}. The mass parameter $M$ describes the inversion of bands with opposite parities at which $M_{0}>0$ correspond to the normal band ordering and $M_{0}<0$ to an inverted one~\cite{w8,w16}.

An important quantity of $\hat{H}_{0}$ is the parameter $\alpha$, which describes the hybridization of the topological surface states with the "parasitic" bulk flat band. An exact $\alpha$ value for a given system can by obtained by \emph{k$\cdot$p} perturbation theory up to linear-in-$k$ order developed in the vicinity of critical points of the Brillouin zone considering all point group symmetries of the bulk crystal. For instance, the Hamiltonian in Eq.~(\ref{eq:1}) at $\alpha=\pi/3$ is essentially the 6-band Kane Hamiltonian regarding the $\Gamma_6$ and $\Gamma_8$ bands, which describes the band structure in the vicinity of the $\Gamma$ point of zinc-blende crystals (see Appendix). This means that, depending on $\alpha$, the Hamiltonian in Eq.~(\ref{eq:1}) interpolates between 3D BHZ Hamiltonian with decoupled flat band at $\alpha=0$ for Bi$_2$Se$_3$-class materials~\cite{w16} and the Kane Hamiltonian at $\alpha=\pi/3$.

Additionally, one can see that $\hat{H}$ in Eq.~(\ref{eq:1}) with $M_{0}=S_{0}=0$, $\hat{k}_{z}=0$ and $\alpha=\pi/4$ corresponds, up to a simple unitary transformation, to the 2D pseudospin-1 Dirac-Weyl Hamiltonian for massless fermions (cf. Refs~\cite{w28,w29}). We note that $\hat{H}(\alpha)$,  $\hat{H}(-\alpha)$ and  $\hat{H}(\pi/2\pm\alpha)$ are all related by unitary transformation.
Also there are no quadratic terms considered in $\hat{H}_{z}$ and $\hat{H}_{0}$ for Eqs.~(\ref{eq:2}) and (\ref{eq:3}) as their form strongly depends on the crystalline symmetry and may differ for two crystals with different values of $\alpha$. Therefore, the universality of the model cannot be preserved beyond the linear approximation but it is in good agreement with magnetooptical experiments for real crystals with $\alpha=0$~\cite{B1,B2} and $\alpha=\pi/3$~\cite{w30,w31,w33,w34}.

Supplementary one can see that $\hat{H}$ in Eq.~(\ref{eq:1}) is invariant under inversion symmetry, but a real crystal may not necessarily feature an inversion center in the unit cell, which can results in additional terms in the Hamiltonian. An explicit form of these terms also depends on the crystalline symmetry and may differ for two crystals with the same value of $\alpha$. For instance, breaking the inversion symmetry in compressively strained HgTe and unstrained Cd$_3$As$_2$, both represented by $\alpha=\pi/3$, results in a transition from Dirac- into Weyl-semimetal in HgTe~\cite{w36a} while for Cd$_3$As$_2$ it retains a fourfold degenerate Dirac node~\cite{w32}. In both crystals, the strength of these terms extracted from experimental data is small~\cite{w30,w31,w33,w34}, and can be neglected. Having said that, we will retain the Hamiltonian in a general form for the reasons of universality and explicitly consider the inversion symmetrical case.

Under these assumptions the Hamiltonian in Eq.~(\ref{eq:1}) has three eigenvalues, each double degenerate due to the time-reversal symmetry. The eigenvalues $E$ follow the equation:
\begin{multline}
\label{eq:4}
\hbar^2v^2(k_x^2+k_y^2)(E_{hh}\cos^2\alpha+E_{lh}\sin^2\alpha)+ \\
+\hbar^2v^2k_z^2E_{hh}=E_cE_{hh}E_{lh},
\end{multline}
where $k_x$, $k_y$, $k_z$ are the quantum numbers of the momentum operators, $E_c(E)=C_{0}+M_{0}-E$, $E_{hh}(E)=C_{0}+S_{0}-E$ and $E_{lh}(E)=C_{0}-M_{0}-E$.

\section{\label{sec:III} Topological surface states}
For the dispersion of the topological surface states, we further consider an interface between two semiconductors with conventional ($M_0^{(I)}>0$, CdTe) and inverted ($M_0^{(II)}<0$, HgTe) band structure. Since the position of all the bands can be different on both sides of the interface it is justified to make the parameters $C_0$ and $S_0$, in addition to $M_0$, dependent on the coordinates. For coordinates far away from the interface the values of these parameters naturally tend to the values inherent to the bulk materials. However, the concrete form of $M_0(x,y,z)$, $C_0(x,y,z)$ and $S_0(x,y,z)$ depends on the smoothness of the junction and the crystallographic orientation.

We will consider two different cases, corresponding to the abrupt junction oriented along different crystallographic directions. As for the other parameters, we consider $v$ and $\alpha$ to be independent of coordinates with the same values from both side of the junction, like it is sufficient for the boundary between CdTe and HgTe~\cite{w31}. Under these assumptions it can be seen from Eq.~(\ref{eq:1}), that the Hamiltonian remains Hermitian even in the presence of the junction, presuming the operators $\hat{k}_x$, $\hat{k}_y$ and $\hat{k}_z$ do not commute.

The initial Schr\"{o}dinger equation with the $6\times6$ Hamiltonian $\hat{H}$ can be considered as a set of differential equations for the $(\varphi_1,\varphi_2,\varphi_3,\varphi_4,\varphi_5,\varphi_6)^{T}$ envelope functions, resulting in an 6$\times$6 differential matrix. First, we note that each pair of $(\varphi_1,\varphi_4)$, $(\varphi_2,\varphi_5)$ and $(\varphi_3,\varphi_6)$ represents the electron states with opposite spin orientation in the given band. Namely, for $\alpha=\pi/3$ in HgTe-class materials, they correspond to the $|\Gamma_6,\pm1/2\rangle$, $|\Gamma_8,\pm3/2\rangle$ and $|\Gamma_8,\pm1/2\rangle$ band, respectively (see Appendix). Second, the absence of $\textbf{k}$-dependent terms in the diagonal elements allows one to express four of the six envelope functions by the two other functions for the same band. Such procedure, which is known as Gaussian elimination, is often used for multi-band Hamiltonians~\cite{w2x2a,w2x2b,w2x2c,w2x2d}.

Considering the pair of $(\varphi_1,\varphi_4)$ envelope functions we can express the four functions $\varphi_2$, $\varphi_3$, $\varphi_5$, $\varphi_6$ in terms of $\varphi_1$ and $\varphi_4$ by keeping the right order of non-commuting operators. Then, substituting the expressions for $\varphi_2$, $\varphi_3$, $\varphi_5$, $\varphi_6$ into the other two equations, we obtain the $2\times2$ energy-dependent Hamiltonian $\hat{H}_E$, which describes the evolution of the vector $\mathbf{\Phi}=(\Phi_1,\Phi_2)^T=(\varphi_1,\varphi_4)^T$ for two spin states:
\begin{equation}
\label{eq:6}
\begin{pmatrix}
\hat{A}+i\hat{B}-E & \hat{K} \\ \hat{K}^{\dag} & \hat{A}-i\hat{B}-E\end{pmatrix}\begin{pmatrix} \Phi_1(x,y,z) \\ \Phi_2(x,y,z) \end{pmatrix}=0,
\end{equation}
where $E$ is the eigenvalue, and
\begin{eqnarray}
\label{eq:7}
\hat{A}=E_c-\hbar^2v^2\hat{k}_z\dfrac{1}{E_{lh}}\hat{k}_z-\hbar^2v^2\hat{k}_x\left(\dfrac{\cos^2\alpha}{E_{lh}}+\dfrac{\sin^2\alpha}{E_{hh}}\right)\hat{k}_x-  \nonumber\\
-\hbar^2v^2\hat{k}_y\left(\dfrac{\cos^2\alpha}{E_{lh}}+\dfrac{\sin^2\alpha}{E_{hh}}\right)\hat{k}_y,~~~~~~
\end{eqnarray}
\begin{eqnarray}
\label{eq:8}
\hat{B}=\hbar^2v^2\hat{k}_y\left(\dfrac{\cos^2\alpha}{E_{lh}}-\dfrac{\sin^2\alpha}{E_{hh}}\right)\hat{k}_x-  ~~~~~~~~~~~~~~~~~~~~~~~~~\nonumber\\
-\hbar^2v^2\hat{k}_x\left(\dfrac{\cos^2\alpha}{E_{lh}}-\dfrac{\sin^2\alpha}{E_{hh}}\right)\hat{k}_y,~~~~~~~~~
\end{eqnarray}
\begin{equation}
\label{eq:9}
\hat{K}=\hbar^2v^2\hat{k}_z\dfrac{\cos\alpha}{E_{lh}}\hat{k}_{-}-\hbar^2v^2\hat{k}_{-}\dfrac{\cos\alpha}{E_{lh}}\hat{k}_z.
\end{equation}\\
Here, we kept the previous notations, and $E_c$, $E_{hh}$ and $E_{lh}$ are the same as for Eq.~(\ref{eq:4}). This changes in the following where they also become dependent on the coordinates:
\begin{eqnarray*}
E_c=C_{0}(x,y,z)+M_{0}(x,y,z)-E,\nonumber\\
E_{hh}=C_{0}(x,y,z)+S_{0}(x,y,z)-E,\nonumber\\
E_{lh}=C_{0}(x,y,z)-M_{0}(x,y,z)-E.
\end{eqnarray*}

We note that the eigenvalue problem in Eq.~(\ref{eq:6}) is very similar to conventional Schr\"{o}dinger equation with the single-band Hamiltonian including non-diagonal spin-orbit interaction~\cite{w2x2a,w2x2b,w2x2c}. Moreover, the procedure described above can be also performed for the other pairs $(\varphi_2,\varphi_5)$ and $(\varphi_3,\varphi_6)$ of the envelope functions.

Although the $2\times2$ energy-dependent Hamiltonian $\hat{H}_E$ in Eq.~(\ref{eq:6}) is non-Hermitian, its eigenvalues are real and are the same as those of $\hat{H}$ in Eq.~(\ref{eq:1}). Recent progress achieved over the last two decades proves that a consistent quantum mechanics can be also built on non-Hermitian Hamiltonians with $\mathcal{PT}$~symmetry~\cite{NH1,NH2,NH3}, where $\mathcal{P}$ and $\mathcal{T}$ are parity and time-reversal operators, respectively. One can see that $\hat{H}_E$ in Eq.~(\ref{eq:6}) can be presented in the form:
\begin{equation*}
\hat{H}_{\mathrm{2x2}}(\mathbf{k})=d_0(\mathbf{k})\sigma_0+d_x(\mathbf{k})\sigma_x+d_y(\mathbf{k})\sigma_y+id_z(\mathbf{k})\sigma_z,
\end{equation*}
where $\sigma_0$ is a 2$\times$2 unity matrix, $\sigma_x$, $\sigma_y$ and $\sigma_z$ are Pauli matrices, $d_0(\mathbf{k})$, $d_x(\mathbf{k})$, $d_y(\mathbf{k})$, $d_z(\mathbf{k})$ are the real functions of $\textbf{k}$. The Hamiltonian $\hat{H}_{\mathrm{2x2}}(\mathbf{k})$ always has the real eigenvalues guaranteed by $\mathcal{PT}$ symmetry (where $\mathcal{PT}=\sigma_x\mathcal{K}$ and $\mathcal{K}$ is the complex conjugation operator)~\cite{NH1,NH2}. In general, non-Hermiticity arises from the presence of energy or particle exchanges with its environment~\cite{NH4,NH5}. The presence of heterojunction in the crystal induces additional interaction between all the bands from the opposite sides of the boundary, which results in the non-Hermiticity of $\hat{H}_E$ describing the pair $(\varphi_1,\varphi_4)$.


\begin{figure*}
\includegraphics [width=2.00\columnwidth, keepaspectratio] {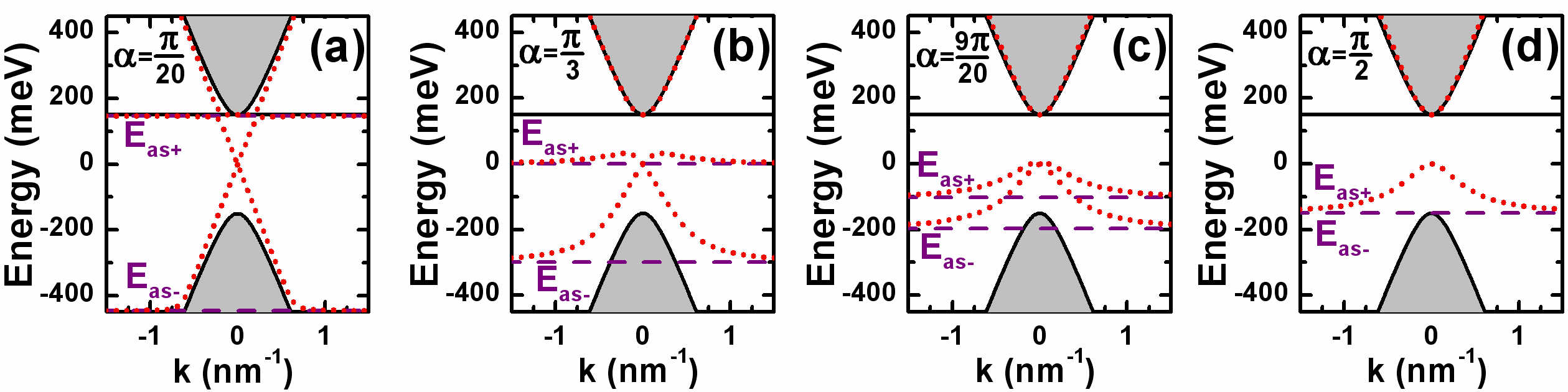} 
\caption{\label{Fig:1} (Color online) Dispersion of the surface states (dotted red) as a function of $k$ (where $k^2=k_x^2+k_y^2$) for $S_0^{(II)}=-M_0^{(II)}$ at different strengths of hybridization with the flat bands: (a) $\alpha=\pi/20$, (b) $\alpha=\pi/3$, (c) $\alpha=9\pi/20$, (d) $\alpha=\pi/2$. The boundary is parallel to the (001) crystallographic plane. The grey regions correspond to the bulk states in the film projected onto the boundary. The solid black curves represent the edges of the projected bulk bands. The bulk states in the barriers arising at $|E|\geq M_0^{(I)}$ are beyond the scale of the panels. The dashed purple lines represent the asymptotic energies $E_{\mathrm{as\pm}}$ at large $k$, given by Eq.~(\ref{eq:16}). The value of $\alpha=\pi/3$ corresponds to unstrained HgCdTe crystals~\cite{w30,w31}.}
\end{figure*}

\subsection{\label{sec:IIIa} Surface states for the boundary parallel to (001) crystallographic plane}
Let us now consider an abrupt semi-infinite boundary parallel to the (001) crystallographic plane, placed at $z=0$. In this case, $M_0(x,y,z)$, $C_0(x,y,z)$ and $S_0(x,y,z)$ only depend on $z$:
\begin{eqnarray}
\label{eq:BC100}
M_0(z)=M_0^{(I)}+\left(M_0^{(II)}-M_0^{(I)}\right)\theta(z),\nonumber\\
C_0(z)=C_0^{(I)}+\left(C_0^{(II)}-C_0^{(I)}\right)\theta(z),\nonumber\\
S_0(z)=S_0^{(I)}+\left(S_0^{(II)}-S_0^{(I)}\right)\theta(z),
\end{eqnarray}
where $\theta(z)$ is a step-like function defined as $\theta(z)=0$ at $z<0$ and $\theta(z)=1$ at $z\geq0$. As mentioned above, the region I corresponds to CdTe with $M_0^{(I)}>0$, while the region II represents HgTe with $M_0^{(II)}<0$. We demonstrate in the following, that the step-like interface allows us to calculate the dispersion of surface states analytically at arbitrary values of $\alpha$. However, the analytical solution can be found at $\alpha=0$ for several smooth interfaces as well~\cite{VPStates}.

The step-like form of $M_0(z)$, $C_0(z)$ and $S_0(z)$ in Eq.~(\ref{eq:BC100}) does not only preserve translation symmetry along the $x$ and $y$ directions but also facilitates the reduction of the eigenvalue problem in Eq.~(\ref{eq:6}) to a set of homogeneous differential equations for the regions $z<0$ and $z>0$. Then the dispersion of the surface states can be found by applying the boundary conditions at $z=0$, which were obtained after integrating Eq.~(\ref{eq:6}) across the small region in the vicinity of $z=0$. Based on the above arguments, $k_x$ and $k_y$ are the good quantum numbers and the wave-function of the surface states localized in the vicinity of $z=0$ has the form:
\begin{eqnarray}
\label{eq:BC100a}
\Phi_{1,2}^{(II)}\sim\exp\left(-\lambda_{z}^{(II)}z\right)\exp\left(ik_xx+ik_yy\right)~
\textnormal{for}~z>0,~~~\nonumber\\
\Phi_{1,2}^{(I)}\sim\exp\left(\lambda_{z}^{(I)}z\right)\exp\left(ik_xx+ik_yy\right),~ \textnormal{for}~z<0,~~~~~~
\end{eqnarray}
where $\lambda_{z}^{(I)}$ and $\lambda_{z}^{(II)}$ are written as:
\begin{equation}
\label{eq:10}
\lambda_{z}^{(n)}=\sqrt{k^2\left(1+\dfrac{E_{lh}^{(n)}-E_{hh}^{(n)}}{E_{hh}^{(n)}}\sin^2\alpha\right)-\dfrac{E_{c}^{(n)}E_{lh}^{(n)}}{\hbar^2v^2}}.
\end{equation}
Here, $n=I,II$ and $k^2=k_x^2+k_y^2$. Note that Eq.~(\ref{eq:10}) can be also derived from Eq.~(\ref{eq:4}) by formally substituting $-k_z^2\rightarrow\lambda_{z}^2$. One can show that the following functions should be continuous across the junction:
\begin{equation}
\label{eq:11}
\begin{pmatrix} \Phi_1 \\[5pt] \Phi_2 \end{pmatrix},~~~~\begin{pmatrix}
\dfrac{1}{E_{lh}(E)}\dfrac{\partial}{\partial z} & -ik_{-}\dfrac{\cos\alpha}{E_{lh}(E)} \\[9pt] ik_{+}\dfrac{\cos\alpha}{E_{lh}(E)} & \dfrac{1}{E_{lh}(E)}\dfrac{\partial}{\partial z}\end{pmatrix}\begin{pmatrix} \Phi_1 \\[5pt]  \Phi_2 \end{pmatrix}.
\end{equation}

Applying the boundary conditions to $\Phi_{1,2}^{(I)}$ and $\Phi_{1,2}^{(II)}$, the secular equation for the non-trivial solution leads to
\begin{equation}
\label{eq:12}
\left(\lambda_{z}^{(I)}E_{lh}^{(II)}+\lambda_{z}^{(II)}E_{lh}^{(I)}\right)^2=
\left(E_{lh}^{(II)}-E_{lh}^{(I)}\right)^2k^2\cos^2\alpha.
\end{equation}
Equations (\ref{eq:10}) and (\ref{eq:12}) give the energy dispersion relations of the surface states. It is seen that the surface states at $k=0$ exist if $E_{lh}^{(I)}$ and $E_{lh}^{(II)}$ are of different sign, which requires different signs of $M_0^{(I)}$ and $M_0^{(II)}$. Substitution of Eq.~(\ref{eq:10}) into Eq.~(\ref{eq:12}) results in a biquadratic equation, which can be solved analytically.

For simplicity, we set $C_0^{(n)}=0$ and analyze the case $S_0^{(n)}=-M_0^{(n)}$. The latter for instance corresponds to the Kane fermions in unstrained HgCdTe crystals~\cite{w30,w31}. In this case, Eq.~(\ref{eq:4}) provides an energy dispersion for the bulk states, which is independent of $\alpha$:
\begin{equation}
\label{eq:5}
\begin{split}
E=C_0\pm\sqrt{M_0^2+\hbar^2v^2(k_x^2+k_y^2)+\hbar^2v^2k_z^2}, \\
E=C_0-M_0,
\end{split}
\end{equation}
while Eq.~(\ref{eq:12}) for the surface states is reduced to
\begin{equation}
\label{eq:13}
A\left(\hbar^2v^2k^2\right)^2-B\hbar^2v^2k^2+E^2(E+M_0^{(I)})^2(E+M_0^{(II)})^2=0,
\end{equation}
where
\begin{multline}
A=\dfrac{(M_0^{(I)}-M_0^{(II)})^2}{4}\sin^4\alpha+ ~~~~~~~~~~~~~~~~~~~~~~~~~~~~~~~~\\ ~~~~~~~~~~~~~~~~~~~+(E+M_0^{(I)})(E+M_0^{(II)})\sin^2\alpha,\\
\label{eq:14}
B=(E^2-M_0^{(I)}M_0^{(II)})(E+M_0^{(I)})(E+M_0^{(II)})\sin^2\alpha+ ~~~~~~~~~~~~~~~~~~~~~~~~~~~~~~\\
+(E+M_0^{(I)})^2(E+M_0^{(II)})^2.
\end{multline}
One can see that for $\alpha=0$, i.e in the absence of hybridization with a flat band, in addition to $E=-M_0^{(I)}$ and $E=-M_0^{(II)}$, Eqs.~(\ref{eq:13}) and (\ref{eq:14}) also give $E=\pm\hbar vk$. The latter coincides with the results obtained within 3D BHZ model with the open boundary conditions~\cite{w41}.

\begin{figure*}
\includegraphics [width=2.0\columnwidth, keepaspectratio] {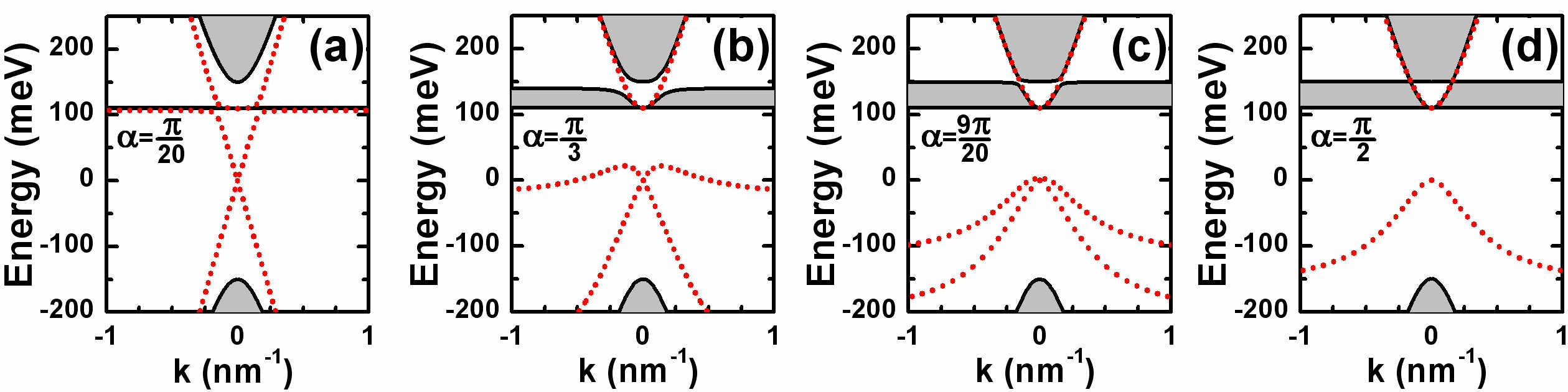} 
\caption{\label{Fig:2} (Color online) Dispersion of the surface states (dotted red) as a function of $k$ (where $k^2=k_x^2+k_y^2$) for $S_0^{(II)}<-M_0^{(II)}$ at different values of $\alpha$. The grey regions correspond to the bulk states in the film projected onto the boundary. The boundary is parallel to (001) crystallographic plane, $S_0^{(II)}=110$~meV. The solid black curves represent the edges of the projected bulk bands. The bulk states in the barriers are beyond the scale of the panels. The value of $\alpha=\pi/3$ corresponds to the tensile strained HgTe crystals.}
\end{figure*}

The presence of hybridization splits initial Dirac-like surface states $E=\pm\hbar vk$ into several branches below and above the flat bands in the materials at both sides of the boundary. Particularly, the surface states lying between the flat bands $E=-M_0^{(I)}$ and $E=-M_0^{(II)}$ are pushed away from the edges of the flat bands at $\alpha\neq0$. At large values of $k$, their asymptotic energies are written as
\begin{equation}
\label{eq:16}
E_{\mathrm{as\pm}}=-\dfrac{M_0^{(I)}+M_0^{(II)}}{2}\pm\dfrac{M_0^{(I)}-M_0^{(II)}}{2}\cos\alpha,
\end{equation}
which can be found from Eqs~(\ref{eq:13}) and~(\ref{eq:14}) with $A=0$. We note that $E_{\mathrm{as\pm}}$ is independent of $k$ only in the linear approximation of $\hat{H}$ in Eq.~(\ref{eq:1}), used for the analytical investigation of the surface states at different values of $\alpha$. Including the quadratic terms, whose explicit form depend on $\alpha$, results in non-zero curvature for $E_{\mathrm{as\pm}}$. The latter has been shown by numerical tight-binding calculations on cubic lattices for the case of $\alpha=\pi/3$, corresponding to real HgTe crystals~\cite{32HgTe}.

Figure~\ref{Fig:1} shows the dispersion of the bulk and surface states for different values of $\alpha$ in the range $(0,\pi/2]$ for an unstrained film with parameters of HgTe ($\hbar v=850$~meV$\cdot$nm~\cite{w31} and $M_0^{(II)}=-150$~meV) sandwiched between CdTe barriers ($M_0^{(I)}=450$~meV). We note that the energy range $-M_0^{(I)}<E<M_0^{(I)}$ corresponds to the band gap in the barriers. Although the bulk dispersion in the film remains the same for any values of $\alpha$ (due to $S_0^{(n)}=-M_0^{(n)}$, see Eq.~(\ref{eq:5})), the dispersion of the surface states strongly depends on hybridization with the flat bands in both materials. At small values of $\alpha$ (see Fig.~\ref{Fig:1}(a)), it consists of four branches $E=E_{\mathrm{as+}}$, $E=\pm\hbar vk$ and $E=E_{\mathrm{as-}}$ anticrossing in the vicinity of the crossing points. In this case, the values of $E_{\mathrm{as+}}$ and $E_{\mathrm{as-}}$ are very close to $-M_0^{(II)}$ and $-M_0^{(I)}$, respectively, since $\cos\alpha\approx1$ in Eq.~(\ref{eq:16}). This picture can be also treated within the conventional degenerate perturbation approach.

\begin{figure*}
\includegraphics [width=2.0\columnwidth, keepaspectratio] {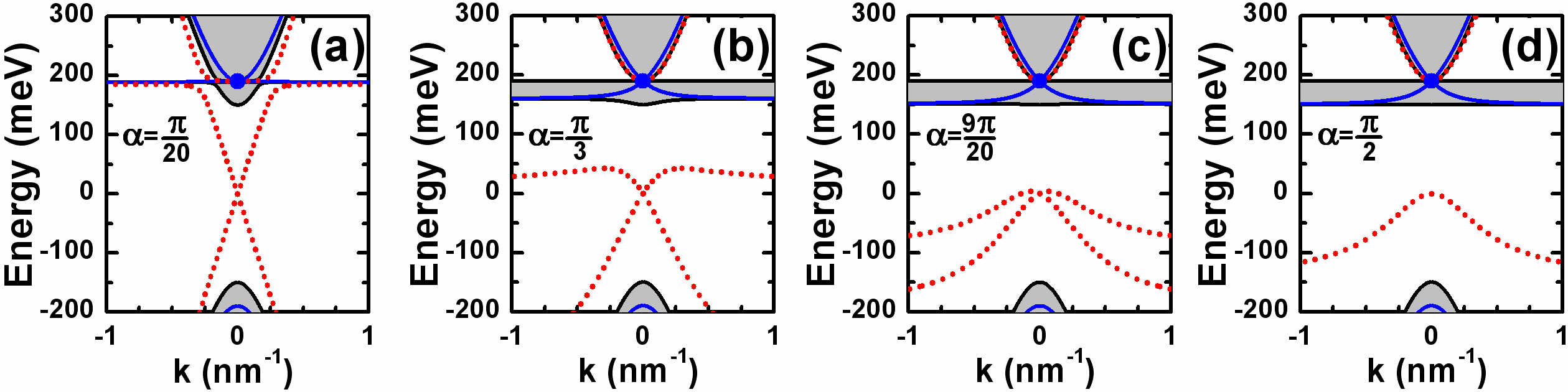} 
\caption{\label{Fig:3} (Color online) Dispersion of the surface states (dotted red) as a function of $k$ (where $k^2=k_x^2+k_y^2$) for $S_0^{(II)}>-M_0^{(II)}$ at different values of $\alpha$. The grey regions correspond to the bulk states in the film projected onto the boundary. The boundary is parallel to (001) crystallographic plane, $S_0^{(II)}=190$~meV. The solid black curves represent the edges of the projected bulk bands. The solid blue curves show the bulk dispersion at $k_z=\pm k_D$, where  $\hbar^2v^2k_D^2=(S_0^{(II)})^2-(M_0^{(II)})^2$. Blue symbols represent projection of bulk Dirac nodes at $\alpha\neq0$ on the boundary surface. The bulk states in the barriers are beyond the scale of the panels. The value of $\alpha=\pi/3$ corresponds to the compressively strained HgTe (or unstrained Cd$_3$As$_2$) crystals.}
\end{figure*}

With increasing of $\alpha$, the surface states for  $-M_0^{(I)}<E<-M_0^{(II)}$ are pushed away from the energies of the flat bulk bands in the materials at both sides of the boundary toward the regions where these bulk states are absent. This is clearly represented by the evolution of the asymptotic energies $E_{\mathrm{as+}}$ and $E_{\mathrm{as-}}$, which are getting closer to each other when $\alpha$ increases as shown in Fig.~\ref{Fig:1}(a-c). Note that dispersion of the surface states remains linear in the vicinity of the $\Gamma$ point of the Brillouin zone. At the specific value of $\alpha=\pi/2$, the asymptotic energies coincide both being equal to $-(M_0^{(I)}+M_0^{(II)})/2$, and the surface states become degenerate, see Fig.~\ref{Fig:1}(d). The latter means the absence of odd-in-$k$ terms in their dispersion. We note that the surface states, similar to those provided in Fig.~\ref{Fig:1}(b) for $\alpha=\pi/3$, were also obtained by more sophisticated numerical calculations based on tight-binding extension of the 6-band Kane Hamiltonian with square terms~\cite{32HgTe}. Although, the results of Ref.~\cite{32HgTe} depend on the constant of artificial cubic lattice used in the calculations, they qualitatively reproduce the dispersion of the surface states at small quasimomentum obtained from our analytical model at $\alpha=\pi/3$.

In addition to the modification of the surface states in the range of $-M_0^{(I)}<E<-M_0^{(II)}$, the hybridization with the flat bulk bands also yields new "massive" branches in the regions above and below the flat bands. We refer to the upper "massive" surface states above the flat band as the \emph{Dyakonov-Khaetskii} (DK) branch. Dyakonov and Khaetskii~\cite{wDK1} were the first, who predicted the massive states at the surface of HgTe crystal. They derived analytically this branch in 1981 by using Luttinger Hamiltonian for the $\Gamma_8$ bands~\cite{wDK2} with an open boundary conditions. In 1985, existence of the localized states at the HgTe/CdTe interface was also predicted for the quantum wells~\cite{wDK3} and superlattices~\cite{wDK4}.

Although the Luttinger Hamiltonian, used in Refs~\cite{wDK1,wDK3}, does not formally consider the inverted $|\Gamma_6,\pm1/2\rangle$ band, this Hamiltonian can be obtained from the 6-band Kane Hamiltonian with the HgTe/CdTe interface by assuming $M_0^{(II)}\rightarrow-\infty$ and $M_0^{(I)}\rightarrow\infty$. Therefore, the upper "massive" surface states in Fig.~\ref{Fig:1} and solution of Dyakonov and Khaetskii~\cite{wDK1}, obtained for particular case of $\alpha=\pi/3$, have the same origin. As seen from Fig.~\ref{Fig:1}, such DK branch is caused by the band inversion in the presence of hybridization with the flat bulk band.

Now we consider a bulk crystal, in which the flat band does not coincide with the bottom of the conduction band, i.e. $S_0^{(II)}\neq-M_0^{(II)}$. The case of $S_0^{(II)}<-M_0^{(II)}$ corresponds to an external tensile biaxial strain, which opens a band gap, yielding 3D TI state~\cite{w23,w24}. The opposite case of $S_0^{(II)}>-M_0^{(II)}$ is realized in the compressively strained HgTe films~\cite{w35a} or in unstrained Cd$_3$As$_2$ crystals~\cite{w33,w34}. As in previous case, we set $C_0^{(n)}=0$ and assume $S_0^{(I)}=-M_0^{(I)}$ in CdTe layer.

Fig.~\ref{Fig:2} shows a picture of the bulk and surface states at different values of $\alpha$ in the $(0,\pi/2]$ range for tensile strained film the with parameters of HgTe ($S_0^{(II)}=110$~meV) sandwiched between CdTe barriers. Note that now the energy dispersions are calculated numerically on the basis of Eq.~(\ref{eq:4}) and Eqs.~(\ref{eq:10}),~(\ref{eq:12}) for the bulk and surface states, respectively. It is seen that energy of the bulk states and the value of a band-gap between the flat and conduction bands strongly depend on $\alpha$. The maximum gap is achieved in the absence of hybridization, while increasing of $\alpha$ leads to a band-gap vanishing. The value of $\alpha=\pi/2$ corresponds to a semimetal with circular nodal line at $k_z=0$ and $k=k_N$, where $k_N^2=2M_0^{(II)}(M_0^{(II)}+S_0^{(II)})/(\hbar^2v^2)$.

As seen from Fig.~\ref{Fig:2}, the surface states in a tensile strained film at different $\alpha$ values remain qualitatively the same, as in Fig.~\ref{Fig:1} for the unstrained film. The main difference is seen in the DK branch, which now exists in the band-gap for the bulk states for all $\alpha$. This is consistent with the general topological arguments claiming that tensile strained HgTe is a 3D TI with gapless surface states~\cite{w23}. However, these surface states can not be represented by massless Dirac fermions, as it is stated in some experimental works on HgTe strained by a CdTe substrate (see, for instance, Refs.~\cite{DF1,DF2,DF3}). Fig.~\ref{Fig:2} clearly shows that the surface states at the HgTe/CdTe boundary of strained HgTe-based 3D TI are "massive" due to the hybridization with heavy-hole $|\Gamma_8,\pm3/2\rangle$ band and represented by DK branch~\cite{wDK1}.

In the opposite case of $S_0^{(II)}>-M_0^{(II)}$, the flat band crosses the conduction band at certain points of the Brillouin zone at $\alpha\neq0$ yielding a 3D Dirac semimetal. At these points, the conduction and flat valence bands can be considered as two highly anisotropic and tilted cones~\cite{w32,w33,w34}, whose nodes lie at $k_x=k_y=0$ and $k_z=\pm k_D$ with $k_D^2=(S_0^2-M_0^2)/(\hbar^2v^2)$, see Eq.~(\ref{eq:4}). Note that at $\alpha=0$, the crossing points are located at the sphere defined by $\hbar^2v^2(k_x^2+k_y^2+k_z^2)=(S_0^2-M_0^2)$, and the 3D Dirac semimetal state does not arise.

Fig.~\ref{Fig:3} presents the bulk and surface states in a film with $S_0^{(II)}>-M_0^{(II)}$ (where $S_0^{(II)}=190$~meV) at different strengths of hybridization with a flat band. As in the previous case of $S_0^{(II)}<-M_0^{(II)}$, the bulk and surface states both depend on the values of $\alpha$.  Interestingly, the dispersion of the surface states in Fig.~\ref{Fig:3} for all $\alpha\neq0$ starts from projection of the bulk Dirac nodes. The bulk band dispersion as a function of $k$ at $k_z=k_D$ is represented by the blue curves. We note that the particular case $\alpha=\pi/3$ is in a good qualitative agreement with the picture of the surface states obtained from the tight-binding calculations for Cd$_3$As$_2$ on a tetragonal lattice~\cite{w32} (see Fig.~3(a,b) therein).

\subsection{\label{sec:IIIb} Surface states for the boundary parallel to (010) plane}
One of the inherent characteristics of the surface states in Dirac semimetals is the existence of a pair of surface Fermi arcs connecting bulk Dirac nodes projected on the surface boundary. The arcs meet at a sharp corner or "kink" at the projected nodes. Such a kink is not allowed in a purely 2D metal, it is a special feature of the crystal symmetry-protected Weyl structure of the Dirac semimetals~\cite{w36}. As the Dirac nodes are located along the (001) crystallographic direction, the surface boundary parallel to the (001) plane has no Fermi arcs.

\begin{figure}
\includegraphics [width=1.0\columnwidth, keepaspectratio] {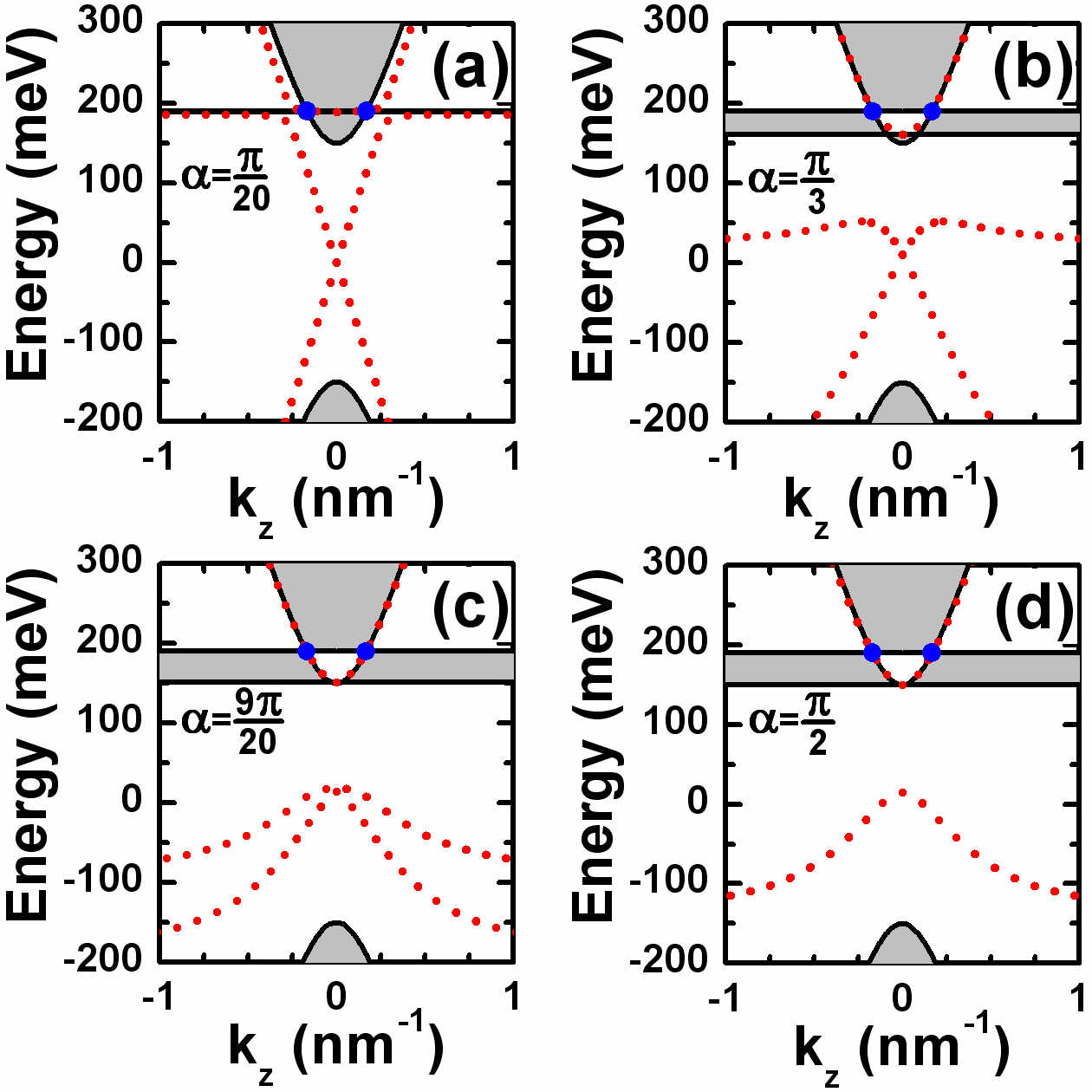} 
\caption{\label{Fig:4} (Color online) Dispersion of the surface (dotted red, $k_x=0$) states as a function of $k_z$ for $S_0^{(II)}>-M_0^{(II)}$ with the boundary parallel to (010) crystallographic plane at different values of $\alpha$. The grey regions correspond to the bulk states in the film projected onto the boundary. The solid black curves represent the edges of the projected bulk bands. As for Fig.~\ref{Fig:3}, we assume $S_0^{(II)}=190$~meV. Blue symbols represent position of bulk Dirac nodes at $\alpha\neq0$. The solid curves marked the edges of the bulk bands. Dispersion of the bulk bands in the barriers is beyond the scale of the panels. The value of $\alpha=\pi/3$ corresponds to the compressively strained HgTe (or unstrained Cd$_3$As$_2$) crystals.}
\end{figure}

Let us now briefly consider the surface states for the boundary containing two projections of the bulk Dirac nodes at $S_0^{(II)}>-M_0^{(II)}$. For the boundary plane parallel to the \emph{x-z} plane and placed at $y=0$, $M_0$, $C_0$ and $S_0$ have a step-like dependence on $y$:
\begin{eqnarray}
\label{eq:BC010}
M_0(y)=M_0^{(I)}+\left(M_0^{(II)}-M_0^{(I)}\right)\theta(y),\nonumber\\
C_0(y)=C_0^{(I)}+\left(C_0^{(II)}-C_0^{(I)}\right)\theta(y),\nonumber\\
S_0(y)=S_0^{(I)}+\left(S_0^{(II)}-S_0^{(I)}\right)\theta(y).
\end{eqnarray}

Basing on the arguments similar to the case of (001) interface, the wave-function of the surface states for the (010) boundary has the form:
\begin{eqnarray}
\label{eq:BC010a}
\Phi_{1,2}^{(II)}\sim\exp\left(-\lambda_{y}^{(II)}y\right)\exp\left(ik_xx+ik_zz\right)~
\textnormal{for}~y>0,~~~\nonumber\\
\Phi_{1,2}^{(I)}\sim\exp\left(\lambda_{y}^{(I)}y\right)\exp\left(ik_xx+ik_zz\right),~ \textnormal{for}~y<0,~~~~~~
\end{eqnarray}
where $\lambda_{y}^{(I)}$ and $\lambda_{y}^{(II)}$ are written as:
\begin{equation}
\label{eq:17}
\lambda_{y}^{(n)}=\sqrt{k_x^2+\dfrac{k_z^2-\frac{E_{c}^{(n)}E_{lh}^{(n)}}{\hbar^2v^2}}{1+\frac{E_{lh}^{(n)}-E_{hh}^{(n)}}{E_{hh}^{(n)}}\sin^2\alpha}},
\end{equation}
where the index $n=I,II$ corresponds to $y<0$ and $y>0$, respectively.

Integration of Eq.~(\ref{eq:6}) across the small region of $y=0$ gives the continuity function across the junction:
\begin{equation}
\label{eq:18}
\begin{pmatrix} \Phi_1 \\[5pt] \Phi_2 \end{pmatrix},~~~~~\begin{pmatrix}
R_{+}\dfrac{\partial}{\partial y}+R_{-}k_x &
k_{z}\dfrac{\cos\alpha}{E_{lh}} \\[9pt]
k_{z}\dfrac{\cos\alpha}{E_{lh}} &
R_{+}\dfrac{\partial}{\partial y}-R_{-}k_x\end{pmatrix}
\begin{pmatrix} \Phi_1 \\[5pt]  \Phi_2 \end{pmatrix},
\end{equation}
where $R_{+}$ and $R_{-}$ are function of $E$ and $\alpha$:
\begin{equation}
\label{eq:19}
R_{\pm}(E)=\dfrac{\cos^2\alpha}{E_{lh}(E)}\pm\dfrac{\sin^2\alpha}{E_{hh}(E)}.
\end{equation}
Applying these boundary conditions to $\Phi_{1,2}^{(I)}$ and $\Phi_{1,2}^{(II)}$, the secular equation for the non-trivial solution leads to
\begin{multline}
\label{eq:20}
\left(\lambda_{y}^{(I)}R_{+}^{(I)}(E)+\lambda_{y}^{(II)}R_{+}^{(II)}(E)\right)^2= \\
=k_z^2\left(\dfrac{1}{E_{lh}^{(II)}}-\dfrac{1}{E_{lh}^{(I)}}\right)^2\cos^2\alpha+ \\
+k_x^2\left(R_{-}^{(II)}(E)-R_{-}^{(I)}(E)\right)^2.
\end{multline}
Substitution of $\lambda_{y}^{(I)}$ and $\lambda_{y}^{(II)}$ into Eq.~(\ref{eq:12}) also results in a quadratic equation for $k_z^2$ and $k_x^2$, which can be solved analytically. Such a quadratic equation gives the energy dispersion and energy contours of the surface states for the boundary parallel to the (010) crystallographic plane.

\begin{figure}
\includegraphics [width=1.0\columnwidth, keepaspectratio] {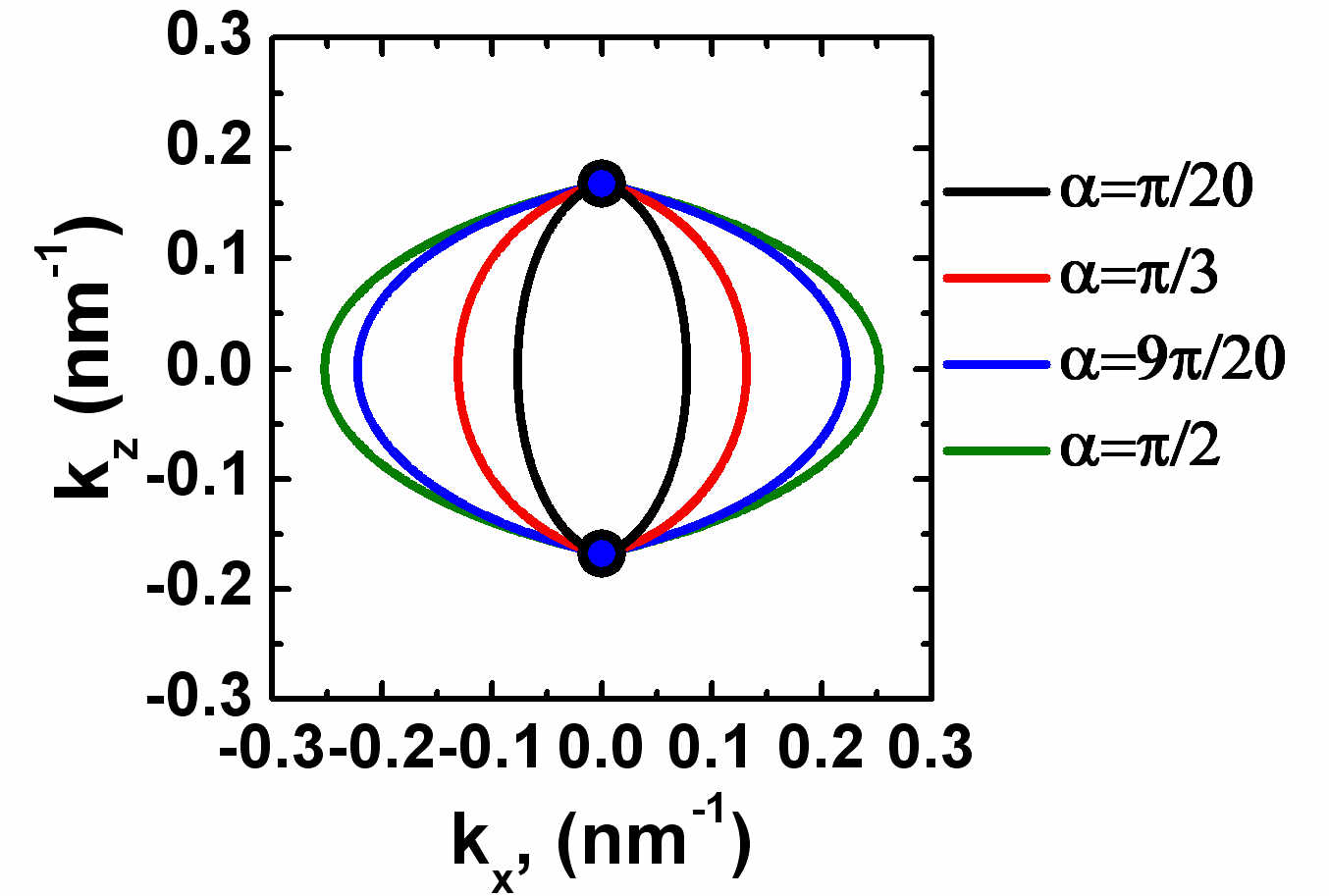} 
\caption{\label{Fig:5} (Color online) Energy contour for the surface states at $E=S_0^{(II)}$ for the boundary parallel to (010) crystallographic plane at $S_0^{(II)}>-M_0^{(II)}$. Blue symbols represent position of bulk Dirac nodes at $\alpha\neq0$. Note that Dirac nodes and Fermi arcs do not exist at $\alpha=0$. The value of $\alpha=\pi/3$ corresponds to the compressively strained HgTe (or unstrained Cd$_3$As$_2$) with crystals.}
\end{figure}

Fig.~\ref{Fig:4} shows the dispersions of the bulk and surface states as a function of $k_z$ for the bulk film with $S_0^{(II)}>-M_0^{(II)}$ and (010) surface boundary. Here, we set $C_0^{(n)}=0$, $S_0^{(I)}=-M_0^{(I)}$, $S_0^{(II)}=190$~meV and assume $k_x=0$. As for the (001) boundary, the picture of surface states for all $\alpha$ values consists of two branches above and below the bottom of the conduction band at $E=-M_0^{(II)}$. As it is seen, the upper DK branch for all $\alpha\neq0$ crosses the bulk dispersion precisely at the Dirac nodes. This stems from the fact that two separated Dirac nodes are connected by the topological surface states~\cite{w36}. A current picture of the surface states for $\alpha=\pi/3$ is also consistent with the tight-binding calculations for Cd$_3$As$_2$~\cite{w32}.

Fig.~\ref{Fig:5} provides energy contours for the surface states at $E=S_0^{(II)}$ at different strengths of hybridization with the flat band. In contrast to the (001) boundary, for which the energy contour of the surface states at the Dirac nodes reduces to a point, the nontrivial surface states at the (010) boundary are clearly visible. Its Fermi surface at $E=S_0^{(II)}$ is composed of two Fermi arcs with the kinks at the projected bulk Dirac nodes. As seen from Fig.~\ref{Fig:5}, the length of Fermi arcs depends on the values of $\alpha$. This means that the period of quantum oscillations originated from cyclotron orbits weaving together Fermi arcs and chiral bulk states~\cite{w36}, should also depend on the hybridization strength.

\section{\label{sec:IV}Summary}
In conclusion, we have performed an analytical study of the hybridization between topological surface states and the non-topological flat band in the bulk. It was shown that the hybridization with the flat band divides the initially Dirac-like surface states, derived from the 3D BHZ model, into two branches, one below the flat band and another one above the edge of conduction band. The upper branch at $\alpha=\pi/3$ is formed by \emph{Dyakonov-Khaetskii} surface states~\cite{wDK1} known for HgTe since the 1980s. Adjusting the hybridization strength, we have explored the evolution of topological surface states in 3D TIs and 3D Dirac semimetals arising at different positions of the flat band. Our results show that the surface states lying inside the band gap of 3D TIs, as well as the Fermi arcs of 3D Dirac semimetals in HgTe and Cd$_3$As$_2$ are represented by the DK branch of the surface states. Although, we have applied the linear approximation for the bulk Hamiltonian, our model qualitatively represents the picture of surface states at small values of $\textbf{k}$, known for $\alpha=\pi/3$ from numerical tight-binding calculations on tetragonal~\cite{w32} and cubic lattices~\cite{32HgTe}. This work paves the way for further analytical investigations of different characteristics of the surface states hybridized with non-topological bands.

\begin{acknowledgments}
The authors gratefully thank S. Gebert (Institut d'Electronique et des Systemes, Montpellier) for the helpful discussions and critical comments. This work was supported by MIPS department of Montpellier University through the "Occitanie Terahertz Platform", by the CNRS through LIA "TeraMIR" and the French Agence Nationale pour la Recherche (Dirac3D project).
\end{acknowledgments}

\appendix
\section*{Appendix: 6-band Kane Hamiltonian}
In order to demonstrate that the Hamiltonian in Eq.~(\ref{eq:1}) for $\alpha=\pi/3$ corresponds to the 6-band Kane Hamiltonian, we consider the 8-band Kane Hamiltonian~\cite{w32o}, whose form is dependent on the choice of the basis set of the Bloch amplitudes for the $\Gamma_6$, $\Gamma_8$ and $\Gamma_7$ bands. In the given basis set
\begin{equation*}
  U_1(\textbf{r})=|\Gamma_6,+1/2\rangle = S\uparrow;
\end{equation*}
\begin{equation*}
  U_2(\textbf{r})=|\Gamma_8,+3/2\rangle = -\frac{i\sqrt{2}}{2}X\uparrow-\frac{i\sqrt{6}}{6}Y\uparrow+\frac{i\sqrt{3}}{3}Z\downarrow;
\end{equation*}
\begin{equation*}
  U_3(\textbf{r})=|\Gamma_8,+1/2\rangle = \frac{\sqrt{2}}{2}X\uparrow-\frac{\sqrt{6}}{6}Y\uparrow+\frac{\sqrt{3}}{3}Z\downarrow;
\end{equation*}
\begin{equation*}
  U_4(\textbf{r})=|\Gamma_7,+1/2\rangle = \frac{i\sqrt{3}}{3}Z\uparrow-\frac{i\sqrt{6}}{3}Y\downarrow;
\end{equation*}
\begin{equation*}
  U_5(\textbf{r})=|\Gamma_6,-1/2\rangle = S\downarrow;
\end{equation*}
\begin{equation*}
  U_6(\textbf{r})=|\Gamma_8,-3/2\rangle = \frac{i\sqrt{2}}{2}X\downarrow+\frac{i\sqrt{6}}{6}Y\downarrow+\frac{i\sqrt{3}}{3}Z\uparrow;
\end{equation*}
\begin{equation*}
  U_7(\textbf{r})=|\Gamma_8,-1/2\rangle = \frac{\sqrt{2}}{2}X\downarrow-\frac{\sqrt{6}}{6}Y\downarrow-\frac{\sqrt{3}}{3}Z\uparrow;
\end{equation*}
\begin{equation*}
  U_8(\textbf{r})=|\Gamma_7,+1/2\rangle = -\frac{i\sqrt{3}}{3}Z\downarrow-\frac{i\sqrt{6}}{3}Y\uparrow,
\end{equation*}
the 8-band Kane Hamiltonian $\hat{H}_{\textbf{k}}^{8\times 8}$ in the presence of only the linear terms takes the form
\begin{widetext}
\begin{equation}\label{eq:8bandKane}
  \hat{H}_{\textbf{k}}^{8\times 8} = \begin{pmatrix}
    E_c & \frac{\sqrt{2}}{2}P\hat{k}_{+} & \frac{\sqrt{6}}{6}P\hat{k}_{-} &
    -\frac{\sqrt{3}}{3}P\hat{k}_{z} & 0 & 0 &
    \frac{\sqrt{6}}{3}P\hat{k}_{z} & -\frac{\sqrt{3}}{3}P\hat{k}_{-}
    \\
    \frac{\sqrt{2}}{2}P\hat{k}_{-} & E_v' & 0 & 0 & 0 & 0 & 0 & 0
    \\
    \frac{\sqrt{6}}{6}P\hat{k}_{+} & 0 & E_v & 0 & \frac{\sqrt{6}}{3}P\hat{k}_{z} & 0 & 0 & 0
    \\
   -\frac{\sqrt{3}}{3}P\hat{k}_{z} & 0 & 0 & E_v''-\Delta & -\frac{\sqrt{3}}{3}P\hat{k}_{-} & 0 &
    0 & 0
    \\
    0 & 0 & \frac{\sqrt{6}}{3}P\hat{k}_{z} & -\frac{\sqrt{3}}{3}P\hat{k}_{+} & E_c & -\frac{\sqrt{2}}{2}P\hat{k}_{-} & -\frac{\sqrt{6}}{6}P\hat{k}_{+} &
    \frac{\sqrt{3}}{3}P\hat{k}_{z}
    \\
    0 & 0 & 0 & 0 & -\frac{\sqrt{2}}{2}P\hat{k}_{+} & E_v' & 0 & 0
    \\
    \frac{\sqrt{6}}{3}P\hat{k}_{z} & 0 & 0 & 0 & -\frac{\sqrt{6}}{6}P\hat{k}_{-} & 0 & E_v & 0
    \\
    -\frac{\sqrt{3}}{3}P\hat{k}_{+} & 0 & 0 & 0 &
    \frac{\sqrt{3}}{3}P\hat{k}_{z} & 0 & 0 & E_v''-\Delta
  \end{pmatrix}.
\end{equation}
Here, $P$ is the Kane momentum matrix element, $\Delta$ is the spin orbit energy and $E_c$ as well as $E_v$ are the conduction and valence band edges, respectively. We note that the values of $E_v$, $E_v'$ and $E_v''$ differ if there is biaxial strain in the (001) crystallographic plane~\cite{Kane}.

In the limit of large $\Delta$, the Hamiltonian $\hat{H}_{\textbf{k}}^{8\times 8}$ can be easily projected on the subspace, orthogonal to the the split-off $\Gamma_7$ band. As we are not interested in terms quadratic in $\mathbf{k}$,the projection is done by simply eliminating the fourth and the eight row and column of the matrix in Eq.~(\ref{eq:8bandKane}):
\begin{equation}\label{eq:6bandKane}
  \hat{H} = \begin{pmatrix}
    E_c & \frac{\sqrt{2}}{2}P\hat{k}_{+} & \frac{\sqrt{6}}{6}P\hat{k}_{-} &
    0 & 0 & \frac{\sqrt{6}}{3}P\hat{k}_{z}
    \\
    \frac{\sqrt{2}}{2}P\hat{k}_{-} & E_v' & 0 & 0 & 0 & 0
    \\
    \frac{\sqrt{6}}{6}P\hat{k}_{+} & 0 & E_v & \frac{\sqrt{6}}{3}P\hat{k}_{z} & 0 & 0
    \\
    0 & 0 & \frac{\sqrt{6}}{3}P\hat{k}_{z} & E_c & -\frac{\sqrt{2}}{2}P\hat{k}_{-} & -\frac{\sqrt{6}}{6}P\hat{k}_{+}
    \\
    0 & 0 & 0 & -\frac{\sqrt{2}}{2}P\hat{k}_{+} & E_v' & 0
    \\
    \frac{\sqrt{6}}{3}P\hat{k}_{z} & 0 & 0 & -\frac{\sqrt{6}}{6}P\hat{k}_{-} & 0 & E_v
  \end{pmatrix}.
\end{equation}
As it is seen, Eq.~(\ref{eq:6bandKane}) coincides with Eq.~(\ref{eq:1}) at $\alpha=\pi/3$ and $\hbar v_{\|}=\hbar v_{\perp}=\sqrt{6}P/3$.
\end{widetext}


\begin{thebibliography}{58}%
\makeatletter
\providecommand \@ifxundefined [1]{%
 \@ifx{#1\undefined}
}%
\providecommand \@ifnum [1]{%
 \ifnum #1\expandafter \@firstoftwo
 \else \expandafter \@secondoftwo
 \fi
}%
\providecommand \@ifx [1]{%
 \ifx #1\expandafter \@firstoftwo
 \else \expandafter \@secondoftwo
 \fi
}%
\providecommand \natexlab [1]{#1}%
\providecommand \enquote  [1]{``#1''}%
\providecommand \bibnamefont  [1]{#1}%
\providecommand \bibfnamefont [1]{#1}%
\providecommand \citenamefont [1]{#1}%
\providecommand \href@noop [0]{\@secondoftwo}%
\providecommand \href [0]{\begingroup \@sanitize@url \@href}%
\providecommand \@href[1]{\@@startlink{#1}\@@href}%
\providecommand \@@href[1]{\endgroup#1\@@endlink}%
\providecommand \@sanitize@url [0]{\catcode `\\12\catcode `\$12\catcode
  `\&12\catcode `\#12\catcode `\^12\catcode `\_12\catcode `\%12\relax}%
\providecommand \@@startlink[1]{}%
\providecommand \@@endlink[0]{}%
\providecommand \url  [0]{\begingroup\@sanitize@url \@url }%
\providecommand \@url [1]{\endgroup\@href {#1}{\urlprefix }}%
\providecommand \urlprefix  [0]{URL }%
\providecommand \Eprint [0]{\href }%
\providecommand \doibase [0]{http://dx.doi.org/}%
\providecommand \selectlanguage [0]{\@gobble}%
\providecommand \bibinfo  [0]{\@secondoftwo}%
\providecommand \bibfield  [0]{\@secondoftwo}%
\providecommand \translation [1]{[#1]}%
\providecommand \BibitemOpen [0]{}%
\providecommand \bibitemStop [0]{}%
\providecommand \bibitemNoStop [0]{.\EOS\space}%
\providecommand \EOS [0]{\spacefactor3000\relax}%
\providecommand \BibitemShut  [1]{\csname bibitem#1\endcsname}%
\let\auto@bib@innerbib\@empty
\bibitem [{\citenamefont {Kane}\ and\ \citenamefont {Mele}(2005)}]{w1}%
  \BibitemOpen
  \bibfield  {author} {\bibinfo {author} {\bibfnamefont {C.~L.}\ \bibnamefont
  {Kane}}\ and\ \bibinfo {author} {\bibfnamefont {E.~J.}\ \bibnamefont
  {Mele}},\ }\href {\doibase 10.1103/PhysRevLett.95.226801} {\bibfield
  {journal} {\bibinfo  {journal} {Phys. Rev. Lett.}\ }\textbf {\bibinfo
  {volume} {95}},\ \bibinfo {pages} {226801} (\bibinfo {year}
  {2005})}\BibitemShut {NoStop}%
\bibitem [{\citenamefont {Fu}\ \emph {et~al.}(2007)\citenamefont {Fu},
  \citenamefont {Kane},\ and\ \citenamefont {Mele}}]{w3}%
  \BibitemOpen
  \bibfield  {author} {\bibinfo {author} {\bibfnamefont {L.}~\bibnamefont
  {Fu}}, \bibinfo {author} {\bibfnamefont {C.~L.}\ \bibnamefont {Kane}}, \ and\
  \bibinfo {author} {\bibfnamefont {E.~J.}\ \bibnamefont {Mele}},\ }\href
  {\doibase 10.1103/PhysRevLett.98.106803} {\bibfield  {journal} {\bibinfo
  {journal} {Phys. Rev. Lett.}\ }\textbf {\bibinfo {volume} {98}},\ \bibinfo
  {pages} {106803} (\bibinfo {year} {2007})}\BibitemShut {NoStop}%
\bibitem [{\citenamefont {Moore}\ and\ \citenamefont {Balents}(2007)}]{w4}%
  \BibitemOpen
  \bibfield  {author} {\bibinfo {author} {\bibfnamefont {J.~E.}\ \bibnamefont
  {Moore}}\ and\ \bibinfo {author} {\bibfnamefont {L.}~\bibnamefont
  {Balents}},\ }\href {\doibase 10.1103/PhysRevB.75.121306} {\bibfield
  {journal} {\bibinfo  {journal} {Phys. Rev. B}\ }\textbf {\bibinfo {volume}
  {75}},\ \bibinfo {pages} {121306} (\bibinfo {year} {2007})}\BibitemShut
  {NoStop}%
\bibitem [{\citenamefont {Roy}(2009)}]{w5}%
  \BibitemOpen
  \bibfield  {author} {\bibinfo {author} {\bibfnamefont {R.}~\bibnamefont
  {Roy}},\ }\href {\doibase 10.1103/PhysRevB.79.195322} {\bibfield  {journal}
  {\bibinfo  {journal} {Phys. Rev. B}\ }\textbf {\bibinfo {volume} {79}},\
  \bibinfo {pages} {195322} (\bibinfo {year} {2009})}\BibitemShut {NoStop}%
\bibitem [{\citenamefont {Hasan}\ and\ \citenamefont {Kane}(2010)}]{w6}%
  \BibitemOpen
  \bibfield  {author} {\bibinfo {author} {\bibfnamefont {M.~Z.}\ \bibnamefont
  {Hasan}}\ and\ \bibinfo {author} {\bibfnamefont {C.~L.}\ \bibnamefont
  {Kane}},\ }\href {\doibase 10.1103/RevModPhys.82.3045} {\bibfield  {journal}
  {\bibinfo  {journal} {Rev. Mod. Phys.}\ }\textbf {\bibinfo {volume} {82}},\
  \bibinfo {pages} {3045} (\bibinfo {year} {2010})}\BibitemShut {NoStop}%
\bibitem [{\citenamefont {Qi}\ and\ \citenamefont {Zhang}(2011)}]{w7}%
  \BibitemOpen
  \bibfield  {author} {\bibinfo {author} {\bibfnamefont {X.-L.}\ \bibnamefont
  {Qi}}\ and\ \bibinfo {author} {\bibfnamefont {S.-C.}\ \bibnamefont {Zhang}},\
  }\href {\doibase 10.1103/RevModPhys.83.1057} {\bibfield  {journal} {\bibinfo
  {journal} {Rev. Mod. Phys.}\ }\textbf {\bibinfo {volume} {83}},\ \bibinfo
  {pages} {1057} (\bibinfo {year} {2011})}\BibitemShut {NoStop}%
\bibitem [{\citenamefont {Bernevig}\ \emph {et~al.}(2006)\citenamefont
  {Bernevig}, \citenamefont {Hughes},\ and\ \citenamefont {Zhang}}]{w8}%
  \BibitemOpen
  \bibfield  {author} {\bibinfo {author} {\bibfnamefont {B.~A.}\ \bibnamefont
  {Bernevig}}, \bibinfo {author} {\bibfnamefont {T.~L.}\ \bibnamefont
  {Hughes}}, \ and\ \bibinfo {author} {\bibfnamefont {S.-C.}\ \bibnamefont
  {Zhang}},\ }\href {\doibase 10.1126/science.1133734} {\bibfield  {journal}
  {\bibinfo  {journal} {Science}\ }\textbf {\bibinfo {volume} {314}},\ \bibinfo
  {pages} {1757} (\bibinfo {year} {2006})}\BibitemShut {NoStop}%
\bibitem [{\citenamefont {K\"{o}nig}\ \emph {et~al.}(2007)\citenamefont
  {K\"{o}nig}, \citenamefont {Wiedmann}, \citenamefont {Br\"{u}ne},
  \citenamefont {Roth}, \citenamefont {Buhmann}, \citenamefont {Molenkamp},
  \citenamefont {Qi},\ and\ \citenamefont {Zhang}}]{w9}%
  \BibitemOpen
  \bibfield  {author} {\bibinfo {author} {\bibfnamefont {M.}~\bibnamefont
  {K\"{o}nig}}, \bibinfo {author} {\bibfnamefont {S.}~\bibnamefont {Wiedmann}},
  \bibinfo {author} {\bibfnamefont {C.}~\bibnamefont {Br\"{u}ne}}, \bibinfo
  {author} {\bibfnamefont {A.}~\bibnamefont {Roth}}, \bibinfo {author}
  {\bibfnamefont {H.}~\bibnamefont {Buhmann}}, \bibinfo {author} {\bibfnamefont
  {L.~W.}\ \bibnamefont {Molenkamp}}, \bibinfo {author} {\bibfnamefont {X.-L.}\
  \bibnamefont {Qi}}, \ and\ \bibinfo {author} {\bibfnamefont {S.-C.}\
  \bibnamefont {Zhang}},\ }\href {\doibase 10.1126/science.1148047} {\bibfield
  {journal} {\bibinfo  {journal} {Science}\ }\textbf {\bibinfo {volume}
  {318}},\ \bibinfo {pages} {766} (\bibinfo {year} {2007})}\BibitemShut
  {NoStop}%
\bibitem [{\citenamefont {Liu}\ \emph {et~al.}(2008)\citenamefont {Liu},
  \citenamefont {Hughes}, \citenamefont {Qi}, \citenamefont {Wang},\ and\
  \citenamefont {Zhang}}]{w10}%
  \BibitemOpen
  \bibfield  {author} {\bibinfo {author} {\bibfnamefont {C.}~\bibnamefont
  {Liu}}, \bibinfo {author} {\bibfnamefont {T.~L.}\ \bibnamefont {Hughes}},
  \bibinfo {author} {\bibfnamefont {X.-L.}\ \bibnamefont {Qi}}, \bibinfo
  {author} {\bibfnamefont {K.}~\bibnamefont {Wang}}, \ and\ \bibinfo {author}
  {\bibfnamefont {S.-C.}\ \bibnamefont {Zhang}},\ }\href {\doibase
  10.1103/PhysRevLett.100.236601} {\bibfield  {journal} {\bibinfo  {journal}
  {Phys. Rev. Lett.}\ }\textbf {\bibinfo {volume} {100}},\ \bibinfo {pages}
  {236601} (\bibinfo {year} {2008})}\BibitemShut {NoStop}%
\bibitem [{\citenamefont {Krishtopenko}\ and\ \citenamefont
  {Teppe}(2018{\natexlab{a}})}]{w12}%
  \BibitemOpen
  \bibfield  {author} {\bibinfo {author} {\bibfnamefont {S.~S.}\ \bibnamefont
  {Krishtopenko}}\ and\ \bibinfo {author} {\bibfnamefont {F.}~\bibnamefont
  {Teppe}},\ }\href {\doibase 10.1126/sciadv.aap7529} {\bibfield  {journal}
  {\bibinfo  {journal} {Sci. Adv.}\ }\textbf {\bibinfo {volume} {4}},\ \bibinfo
  {pages} {eaap7529} (\bibinfo {year} {2018}{\natexlab{a}})}\BibitemShut
  {NoStop}%
\bibitem [{\citenamefont {Krishtopenko}\ \emph {et~al.}(2018)\citenamefont
  {Krishtopenko}, \citenamefont {Ruffenach}, \citenamefont {Gonzalez-Posada},
  \citenamefont {Boissier}, \citenamefont {Marcinkiewicz}, \citenamefont
  {Fadeev}, \citenamefont {Kadykov}, \citenamefont {Rumyantsev}, \citenamefont
  {Morozov}, \citenamefont {Gavrilenko}, \citenamefont {Consejo}, \citenamefont
  {Desrat}, \citenamefont {Jouault}, \citenamefont {Knap}, \citenamefont
  {Tourni\'e},\ and\ \citenamefont {Teppe}}]{w13}%
  \BibitemOpen
  \bibfield  {author} {\bibinfo {author} {\bibfnamefont {S.~S.}\ \bibnamefont
  {Krishtopenko}}, \bibinfo {author} {\bibfnamefont {S.}~\bibnamefont
  {Ruffenach}}, \bibinfo {author} {\bibfnamefont {F.}~\bibnamefont
  {Gonzalez-Posada}}, \bibinfo {author} {\bibfnamefont {G.}~\bibnamefont
  {Boissier}}, \bibinfo {author} {\bibfnamefont {M.}~\bibnamefont
  {Marcinkiewicz}}, \bibinfo {author} {\bibfnamefont {M.~A.}\ \bibnamefont
  {Fadeev}}, \bibinfo {author} {\bibfnamefont {A.~M.}\ \bibnamefont {Kadykov}},
  \bibinfo {author} {\bibfnamefont {V.~V.}\ \bibnamefont {Rumyantsev}},
  \bibinfo {author} {\bibfnamefont {S.~V.}\ \bibnamefont {Morozov}}, \bibinfo
  {author} {\bibfnamefont {V.~I.}\ \bibnamefont {Gavrilenko}}, \bibinfo
  {author} {\bibfnamefont {C.}~\bibnamefont {Consejo}}, \bibinfo {author}
  {\bibfnamefont {W.}~\bibnamefont {Desrat}}, \bibinfo {author} {\bibfnamefont
  {B.}~\bibnamefont {Jouault}}, \bibinfo {author} {\bibfnamefont
  {W.}~\bibnamefont {Knap}}, \bibinfo {author} {\bibfnamefont {E.}~\bibnamefont
  {Tourni\'e}}, \ and\ \bibinfo {author} {\bibfnamefont {F.}~\bibnamefont
  {Teppe}},\ }\href {\doibase 10.1103/PhysRevB.97.245419} {\bibfield  {journal}
  {\bibinfo  {journal} {Phys. Rev. B}\ }\textbf {\bibinfo {volume} {97}},\
  \bibinfo {pages} {245419} (\bibinfo {year} {2018})}\BibitemShut {NoStop}%
\bibitem [{\citenamefont {Wu}\ \emph {et~al.}(2018)\citenamefont {Wu},
  \citenamefont {Fatemi}, \citenamefont {Gibson}, \citenamefont {Watanabe},
  \citenamefont {Taniguchi}, \citenamefont {Cava},\ and\ \citenamefont
  {Jarillo-Herrero}}]{w14}%
  \BibitemOpen
  \bibfield  {author} {\bibinfo {author} {\bibfnamefont {S.}~\bibnamefont
  {Wu}}, \bibinfo {author} {\bibfnamefont {V.}~\bibnamefont {Fatemi}}, \bibinfo
  {author} {\bibfnamefont {Q.~D.}\ \bibnamefont {Gibson}}, \bibinfo {author}
  {\bibfnamefont {K.}~\bibnamefont {Watanabe}}, \bibinfo {author}
  {\bibfnamefont {T.}~\bibnamefont {Taniguchi}}, \bibinfo {author}
  {\bibfnamefont {R.~J.}\ \bibnamefont {Cava}}, \ and\ \bibinfo {author}
  {\bibfnamefont {P.}~\bibnamefont {Jarillo-Herrero}},\ }\href {\doibase
  10.1126/science.aan6003} {\bibfield  {journal} {\bibinfo  {journal}
  {Science}\ }\textbf {\bibinfo {volume} {359}},\ \bibinfo {pages} {76}
  (\bibinfo {year} {2018})}\BibitemShut {NoStop}%
\bibitem [{\citenamefont {Krishtopenko}\ and\ \citenamefont
  {Teppe}(2018{\natexlab{b}})}]{w26}%
  \BibitemOpen
  \bibfield  {author} {\bibinfo {author} {\bibfnamefont {S.~S.}\ \bibnamefont
  {Krishtopenko}}\ and\ \bibinfo {author} {\bibfnamefont {F.}~\bibnamefont
  {Teppe}},\ }\href {\doibase 10.1103/PhysRevB.97.165408} {\bibfield  {journal}
  {\bibinfo  {journal} {Phys. Rev. B}\ }\textbf {\bibinfo {volume} {97}},\
  \bibinfo {pages} {165408} (\bibinfo {year} {2018}{\natexlab{b}})}\BibitemShut
  {NoStop}%
\bibitem [{\citenamefont {Xia}\ \emph {et~al.}(2009)\citenamefont {Xia},
  \citenamefont {Qian}, \citenamefont {Hsieh}, \citenamefont {Wray},
  \citenamefont {Pal}, \citenamefont {Lin}, \citenamefont {Bansil},
  \citenamefont {Grauer}, \citenamefont {Hor}, \citenamefont {Cava},\ and\
  \citenamefont {Hasan}}]{w15}%
  \BibitemOpen
  \bibfield  {author} {\bibinfo {author} {\bibfnamefont {Y.}~\bibnamefont
  {Xia}}, \bibinfo {author} {\bibfnamefont {D.}~\bibnamefont {Qian}}, \bibinfo
  {author} {\bibfnamefont {D.}~\bibnamefont {Hsieh}}, \bibinfo {author}
  {\bibfnamefont {L.}~\bibnamefont {Wray}}, \bibinfo {author} {\bibfnamefont
  {A.}~\bibnamefont {Pal}}, \bibinfo {author} {\bibfnamefont {H.}~\bibnamefont
  {Lin}}, \bibinfo {author} {\bibfnamefont {A.}~\bibnamefont {Bansil}},
  \bibinfo {author} {\bibfnamefont {D.}~\bibnamefont {Grauer}}, \bibinfo
  {author} {\bibfnamefont {Y.~S.}\ \bibnamefont {Hor}}, \bibinfo {author}
  {\bibfnamefont {R.~J.}\ \bibnamefont {Cava}}, \ and\ \bibinfo {author}
  {\bibfnamefont {M.~Z.}\ \bibnamefont {Hasan}},\ }\href {\doibase
  10.1038/nphys1274} {\bibfield  {journal} {\bibinfo  {journal} {Nature Phys.}\
  }\textbf {\bibinfo {volume} {5}},\ \bibinfo {pages} {398} (\bibinfo {year}
  {2009})}\BibitemShut {NoStop}%
\bibitem [{\citenamefont {Zhang}\ \emph {et~al.}(2009)\citenamefont {Zhang},
  \citenamefont {Liu}, \citenamefont {Qi}, \citenamefont {Dai}, \citenamefont
  {Fang},\ and\ \citenamefont {Zhang}}]{w16}%
  \BibitemOpen
  \bibfield  {author} {\bibinfo {author} {\bibfnamefont {H.}~\bibnamefont
  {Zhang}}, \bibinfo {author} {\bibfnamefont {C.-X.}\ \bibnamefont {Liu}},
  \bibinfo {author} {\bibfnamefont {X.-L.}\ \bibnamefont {Qi}}, \bibinfo
  {author} {\bibfnamefont {X.}~\bibnamefont {Dai}}, \bibinfo {author}
  {\bibfnamefont {Z.}~\bibnamefont {Fang}}, \ and\ \bibinfo {author}
  {\bibfnamefont {S.-C.}\ \bibnamefont {Zhang}},\ }\href {\doibase
  10.1038/nphys1270} {\bibfield  {journal} {\bibinfo  {journal} {Nature Phys.}\
  }\textbf {\bibinfo {volume} {5}},\ \bibinfo {pages} {438} (\bibinfo {year}
  {2009})}\BibitemShut {NoStop}%
\bibitem [{\citenamefont {Chen}\ \emph {et~al.}(2009)\citenamefont {Chen},
  \citenamefont {Analytis}, \citenamefont {Chu}, \citenamefont {Liu},
  \citenamefont {Mo}, \citenamefont {Qi}, \citenamefont {Zhang}, \citenamefont
  {Lu}, \citenamefont {Dai}, \citenamefont {Fang}, \citenamefont {Zhang},
  \citenamefont {Fisher}, \citenamefont {Hussain},\ and\ \citenamefont
  {Shen}}]{w17}%
  \BibitemOpen
  \bibfield  {author} {\bibinfo {author} {\bibfnamefont {Y.~L.}\ \bibnamefont
  {Chen}}, \bibinfo {author} {\bibfnamefont {J.~G.}\ \bibnamefont {Analytis}},
  \bibinfo {author} {\bibfnamefont {J.-H.}\ \bibnamefont {Chu}}, \bibinfo
  {author} {\bibfnamefont {Z.~K.}\ \bibnamefont {Liu}}, \bibinfo {author}
  {\bibfnamefont {S.-K.}\ \bibnamefont {Mo}}, \bibinfo {author} {\bibfnamefont
  {X.~L.}\ \bibnamefont {Qi}}, \bibinfo {author} {\bibfnamefont {H.~J.}\
  \bibnamefont {Zhang}}, \bibinfo {author} {\bibfnamefont {D.~H.}\ \bibnamefont
  {Lu}}, \bibinfo {author} {\bibfnamefont {X.}~\bibnamefont {Dai}}, \bibinfo
  {author} {\bibfnamefont {Z.}~\bibnamefont {Fang}}, \bibinfo {author}
  {\bibfnamefont {S.~C.}\ \bibnamefont {Zhang}}, \bibinfo {author}
  {\bibfnamefont {I.~R.}\ \bibnamefont {Fisher}}, \bibinfo {author}
  {\bibfnamefont {Z.}~\bibnamefont {Hussain}}, \ and\ \bibinfo {author}
  {\bibfnamefont {Z.-X.}\ \bibnamefont {Shen}},\ }\href {\doibase
  10.1126/science.1173034} {\bibfield  {journal} {\bibinfo  {journal}
  {Science}\ }\textbf {\bibinfo {volume} {325}},\ \bibinfo {pages} {178}
  (\bibinfo {year} {2009})}\BibitemShut {NoStop}%
\bibitem [{\citenamefont {Hsieh}\ \emph {et~al.}(2009)\citenamefont {Hsieh},
  \citenamefont {Xia}, \citenamefont {Wray}, \citenamefont {Qian},
  \citenamefont {Pal}, \citenamefont {Dil}, \citenamefont {Osterwalder},
  \citenamefont {Meier}, \citenamefont {Bihlmayer}, \citenamefont {Kane},
  \citenamefont {Hor}, \citenamefont {Cava},\ and\ \citenamefont
  {Hasan}}]{w18}%
  \BibitemOpen
  \bibfield  {author} {\bibinfo {author} {\bibfnamefont {D.}~\bibnamefont
  {Hsieh}}, \bibinfo {author} {\bibfnamefont {Y.}~\bibnamefont {Xia}}, \bibinfo
  {author} {\bibfnamefont {L.}~\bibnamefont {Wray}}, \bibinfo {author}
  {\bibfnamefont {D.}~\bibnamefont {Qian}}, \bibinfo {author} {\bibfnamefont
  {A.}~\bibnamefont {Pal}}, \bibinfo {author} {\bibfnamefont {J.~H.}\
  \bibnamefont {Dil}}, \bibinfo {author} {\bibfnamefont {J.}~\bibnamefont
  {Osterwalder}}, \bibinfo {author} {\bibfnamefont {F.}~\bibnamefont {Meier}},
  \bibinfo {author} {\bibfnamefont {G.}~\bibnamefont {Bihlmayer}}, \bibinfo
  {author} {\bibfnamefont {C.~L.}\ \bibnamefont {Kane}}, \bibinfo {author}
  {\bibfnamefont {Y.~S.}\ \bibnamefont {Hor}}, \bibinfo {author} {\bibfnamefont
  {R.~J.}\ \bibnamefont {Cava}}, \ and\ \bibinfo {author} {\bibfnamefont
  {M.~Z.}\ \bibnamefont {Hasan}},\ }\href {\doibase 10.1126/science.1167733}
  {\bibfield  {journal} {\bibinfo  {journal} {Science}\ }\textbf {\bibinfo
  {volume} {323}},\ \bibinfo {pages} {919} (\bibinfo {year}
  {2009})}\BibitemShut {NoStop}%
\bibitem [{\citenamefont {Fu}(2011)}]{w19}%
  \BibitemOpen
  \bibfield  {author} {\bibinfo {author} {\bibfnamefont {L.}~\bibnamefont
  {Fu}},\ }\href {\doibase 10.1103/PhysRevLett.106.106802} {\bibfield
  {journal} {\bibinfo  {journal} {Phys. Rev. Lett.}\ }\textbf {\bibinfo
  {volume} {106}},\ \bibinfo {pages} {106802} (\bibinfo {year}
  {2011})}\BibitemShut {NoStop}%
\bibitem [{\citenamefont {Hsieh}\ \emph {et~al.}(2012)\citenamefont {Hsieh},
  \citenamefont {Lin}, \citenamefont {Liu}, \citenamefont {Duan}, \citenamefont
  {Bansil},\ and\ \citenamefont {Fu}}]{w20}%
  \BibitemOpen
  \bibfield  {author} {\bibinfo {author} {\bibfnamefont {T.~H.}\ \bibnamefont
  {Hsieh}}, \bibinfo {author} {\bibfnamefont {H.}~\bibnamefont {Lin}}, \bibinfo
  {author} {\bibfnamefont {J.}~\bibnamefont {Liu}}, \bibinfo {author}
  {\bibfnamefont {W.}~\bibnamefont {Duan}}, \bibinfo {author} {\bibfnamefont
  {A.}~\bibnamefont {Bansil}}, \ and\ \bibinfo {author} {\bibfnamefont
  {L.}~\bibnamefont {Fu}},\ }\href {\doibase 10.1038/ncomms1969} {\bibfield
  {journal} {\bibinfo  {journal} {Nat. Commun.}\ }\textbf {\bibinfo {volume}
  {3}},\ \bibinfo {pages} {982} (\bibinfo {year} {2012})}\BibitemShut {NoStop}%
\bibitem [{\citenamefont {Dziawa}\ \emph {et~al.}(2012)\citenamefont {Dziawa},
  \citenamefont {Kowalski}, \citenamefont {Dybko}, \citenamefont {Buczko},
  \citenamefont {Szczerbakow}, \citenamefont {Szot}, \citenamefont
  {Lusakowska}, \citenamefont {Balasubramanian}, \citenamefont {Wojek},
  \citenamefont {Berntsen}, \citenamefont {Tjernberg},\ and\ \citenamefont
  {Story}}]{w21}%
  \BibitemOpen
  \bibfield  {author} {\bibinfo {author} {\bibfnamefont {P.}~\bibnamefont
  {Dziawa}}, \bibinfo {author} {\bibfnamefont {B.~J.}\ \bibnamefont
  {Kowalski}}, \bibinfo {author} {\bibfnamefont {K.}~\bibnamefont {Dybko}},
  \bibinfo {author} {\bibfnamefont {R.}~\bibnamefont {Buczko}}, \bibinfo
  {author} {\bibfnamefont {A.}~\bibnamefont {Szczerbakow}}, \bibinfo {author}
  {\bibfnamefont {M.}~\bibnamefont {Szot}}, \bibinfo {author} {\bibfnamefont
  {E.}~\bibnamefont {Lusakowska}}, \bibinfo {author} {\bibfnamefont
  {T.}~\bibnamefont {Balasubramanian}}, \bibinfo {author} {\bibfnamefont
  {B.~M.}\ \bibnamefont {Wojek}}, \bibinfo {author} {\bibfnamefont {M.~H.}\
  \bibnamefont {Berntsen}}, \bibinfo {author} {\bibfnamefont {O.}~\bibnamefont
  {Tjernberg}}, \ and\ \bibinfo {author} {\bibfnamefont {T.}~\bibnamefont
  {Story}},\ }\href {\doibase 10.1038/nmat3449} {\bibfield  {journal} {\bibinfo
   {journal} {Nat. Mater.}\ }\textbf {\bibinfo {volume} {11}},\ \bibinfo
  {pages} {1023} (\bibinfo {year} {2012})}\BibitemShut {NoStop}%
\bibitem [{\citenamefont {Okada}\ \emph {et~al.}(2013)\citenamefont {Okada},
  \citenamefont {Serbyn}, \citenamefont {Lin}, \citenamefont {Walkup},
  \citenamefont {Zhou}, \citenamefont {Dhital}, \citenamefont {Neupane},
  \citenamefont {Xu}, \citenamefont {Wang}, \citenamefont {Sankar},
  \citenamefont {Chou}, \citenamefont {Bansil}, \citenamefont {Hasan},
  \citenamefont {Wilson}, \citenamefont {Fu},\ and\ \citenamefont
  {Madhavan}}]{w22}%
  \BibitemOpen
  \bibfield  {author} {\bibinfo {author} {\bibfnamefont {Y.}~\bibnamefont
  {Okada}}, \bibinfo {author} {\bibfnamefont {M.}~\bibnamefont {Serbyn}},
  \bibinfo {author} {\bibfnamefont {H.}~\bibnamefont {Lin}}, \bibinfo {author}
  {\bibfnamefont {D.}~\bibnamefont {Walkup}}, \bibinfo {author} {\bibfnamefont
  {W.}~\bibnamefont {Zhou}}, \bibinfo {author} {\bibfnamefont {C.}~\bibnamefont
  {Dhital}}, \bibinfo {author} {\bibfnamefont {M.}~\bibnamefont {Neupane}},
  \bibinfo {author} {\bibfnamefont {S.}~\bibnamefont {Xu}}, \bibinfo {author}
  {\bibfnamefont {Y.~J.}\ \bibnamefont {Wang}}, \bibinfo {author}
  {\bibfnamefont {R.}~\bibnamefont {Sankar}}, \bibinfo {author} {\bibfnamefont
  {F.}~\bibnamefont {Chou}}, \bibinfo {author} {\bibfnamefont {A.}~\bibnamefont
  {Bansil}}, \bibinfo {author} {\bibfnamefont {M.~Z.}\ \bibnamefont {Hasan}},
  \bibinfo {author} {\bibfnamefont {S.~D.}\ \bibnamefont {Wilson}}, \bibinfo
  {author} {\bibfnamefont {L.}~\bibnamefont {Fu}}, \ and\ \bibinfo {author}
  {\bibfnamefont {V.}~\bibnamefont {Madhavan}},\ }\href {\doibase
  10.1126/science.1239451} {\bibfield  {journal} {\bibinfo  {journal}
  {Science}\ }\textbf {\bibinfo {volume} {341}},\ \bibinfo {pages} {1496}
  (\bibinfo {year} {2013})}\BibitemShut {NoStop}%
\bibitem [{\citenamefont {Fu}\ and\ \citenamefont {Kane}(2007)}]{w23}%
  \BibitemOpen
  \bibfield  {author} {\bibinfo {author} {\bibfnamefont {L.}~\bibnamefont
  {Fu}}\ and\ \bibinfo {author} {\bibfnamefont {C.~L.}\ \bibnamefont {Kane}},\
  }\href {\doibase 10.1103/PhysRevB.76.045302} {\bibfield  {journal} {\bibinfo
  {journal} {Phys. Rev. B}\ }\textbf {\bibinfo {volume} {76}},\ \bibinfo
  {pages} {045302} (\bibinfo {year} {2007})}\BibitemShut {NoStop}%
\bibitem [{\citenamefont {Dai}\ \emph {et~al.}(2008)\citenamefont {Dai},
  \citenamefont {Hughes}, \citenamefont {Qi}, \citenamefont {Fang},\ and\
  \citenamefont {Zhang}}]{w23a}%
  \BibitemOpen
  \bibfield  {author} {\bibinfo {author} {\bibfnamefont {X.}~\bibnamefont
  {Dai}}, \bibinfo {author} {\bibfnamefont {T.~L.}\ \bibnamefont {Hughes}},
  \bibinfo {author} {\bibfnamefont {X.-L.}\ \bibnamefont {Qi}}, \bibinfo
  {author} {\bibfnamefont {Z.}~\bibnamefont {Fang}}, \ and\ \bibinfo {author}
  {\bibfnamefont {S.-C.}\ \bibnamefont {Zhang}},\ }\href {\doibase
  10.1103/PhysRevB.77.125319} {\bibfield  {journal} {\bibinfo  {journal} {Phys.
  Rev. B}\ }\textbf {\bibinfo {volume} {77}},\ \bibinfo {pages} {125319}
  (\bibinfo {year} {2008})}\BibitemShut {NoStop}%
\bibitem [{\citenamefont {Br\"une}\ \emph {et~al.}(2011)\citenamefont
  {Br\"une}, \citenamefont {Liu}, \citenamefont {Novik}, \citenamefont
  {Hankiewicz}, \citenamefont {Buhmann}, \citenamefont {Chen}, \citenamefont
  {Qi}, \citenamefont {Shen}, \citenamefont {Zhang},\ and\ \citenamefont
  {Molenkamp}}]{w24}%
  \BibitemOpen
  \bibfield  {author} {\bibinfo {author} {\bibfnamefont {C.}~\bibnamefont
  {Br\"une}}, \bibinfo {author} {\bibfnamefont {C.~X.}\ \bibnamefont {Liu}},
  \bibinfo {author} {\bibfnamefont {E.~G.}\ \bibnamefont {Novik}}, \bibinfo
  {author} {\bibfnamefont {E.~M.}\ \bibnamefont {Hankiewicz}}, \bibinfo
  {author} {\bibfnamefont {H.}~\bibnamefont {Buhmann}}, \bibinfo {author}
  {\bibfnamefont {Y.~L.}\ \bibnamefont {Chen}}, \bibinfo {author}
  {\bibfnamefont {X.~L.}\ \bibnamefont {Qi}}, \bibinfo {author} {\bibfnamefont
  {Z.~X.}\ \bibnamefont {Shen}}, \bibinfo {author} {\bibfnamefont {S.~C.}\
  \bibnamefont {Zhang}}, \ and\ \bibinfo {author} {\bibfnamefont {L.~W.}\
  \bibnamefont {Molenkamp}},\ }\href {\doibase 10.1103/PhysRevLett.106.126803}
  {\bibfield  {journal} {\bibinfo  {journal} {Phys. Rev. Lett.}\ }\textbf
  {\bibinfo {volume} {106}},\ \bibinfo {pages} {126803} (\bibinfo {year}
  {2011})}\BibitemShut {NoStop}%
\bibitem [{\citenamefont {Dyakonov}\ and\ \citenamefont
  {Khaetskii}(1981)}]{wDK1}%
  \BibitemOpen
  \bibfield  {author} {\bibinfo {author} {\bibfnamefont {M.~I.}\ \bibnamefont
  {Dyakonov}}\ and\ \bibinfo {author} {\bibfnamefont {A.~V.}\ \bibnamefont
  {Khaetskii}},\ }\href@noop {} {\bibfield  {journal} {\bibinfo  {journal}
  {JETP Lett.}\ }\textbf {\bibinfo {volume} {33}},\ \bibinfo {pages} {110}
  (\bibinfo {year} {1981})}\BibitemShut {NoStop}%
\bibitem [{\citenamefont {Kibis}\ \emph {et~al.}(2019)\citenamefont {Kibis},
  \citenamefont {Kyriienko},\ and\ \citenamefont {Shelykh}}]{wDK1a}%
  \BibitemOpen
  \bibfield  {author} {\bibinfo {author} {\bibfnamefont {O.~V.}\ \bibnamefont
  {Kibis}}, \bibinfo {author} {\bibfnamefont {O.}~\bibnamefont {Kyriienko}}, \
  and\ \bibinfo {author} {\bibfnamefont {I.~A.}\ \bibnamefont {Shelykh}},\
  }\href {\doibase 10.1088/1367-2630/ab1406} {\bibfield  {journal} {\bibinfo
  {journal} {New J. Phys.}\ }\textbf {\bibinfo {volume} {21}},\ \bibinfo
  {pages} {043016} (\bibinfo {year} {2019})}\BibitemShut {NoStop}%
\bibitem [{\citenamefont {Mahler}\ \emph {et~al.}(2019)\citenamefont {Mahler},
  \citenamefont {Mayer}, \citenamefont {Leubner}, \citenamefont {Lunczer},
  \citenamefont {Di~Sante}, \citenamefont {Sangiovanni}, \citenamefont
  {Thomale}, \citenamefont {Hankiewicz}, \citenamefont {Buhmann}, \citenamefont
  {Gould},\ and\ \citenamefont {Molenkamp}}]{w35a}%
  \BibitemOpen
  \bibfield  {author} {\bibinfo {author} {\bibfnamefont {D.~M.}\ \bibnamefont
  {Mahler}}, \bibinfo {author} {\bibfnamefont {J.-B.}\ \bibnamefont {Mayer}},
  \bibinfo {author} {\bibfnamefont {P.}~\bibnamefont {Leubner}}, \bibinfo
  {author} {\bibfnamefont {L.}~\bibnamefont {Lunczer}}, \bibinfo {author}
  {\bibfnamefont {D.}~\bibnamefont {Di~Sante}}, \bibinfo {author}
  {\bibfnamefont {G.}~\bibnamefont {Sangiovanni}}, \bibinfo {author}
  {\bibfnamefont {R.}~\bibnamefont {Thomale}}, \bibinfo {author} {\bibfnamefont
  {E.~M.}\ \bibnamefont {Hankiewicz}}, \bibinfo {author} {\bibfnamefont
  {H.}~\bibnamefont {Buhmann}}, \bibinfo {author} {\bibfnamefont
  {C.}~\bibnamefont {Gould}}, \ and\ \bibinfo {author} {\bibfnamefont {L.~W.}\
  \bibnamefont {Molenkamp}},\ }\href {\doibase 10.1103/PhysRevX.9.031034}
  {\bibfield  {journal} {\bibinfo  {journal} {Phys. Rev. X}\ }\textbf {\bibinfo
  {volume} {9}},\ \bibinfo {pages} {031034} (\bibinfo {year}
  {2019})}\BibitemShut {NoStop}%
\bibitem [{\citenamefont {Bodnar}()}]{w32o}%
  \BibitemOpen
  \bibfield  {author} {\bibinfo {author} {\bibfnamefont {J.}~\bibnamefont
  {Bodnar}},\ }\href {http://arxiv.org/abs/1709.05845} {\bibinfo  {journal}
  {Proc. III Conf. Narrow-Gap Semiconductors, Warsaw, edited by J.
  Rauluszkiewicz, M. G\'{o}rska, and E. Kaczmarek (Elsevier, 1977) pp. 311, see
  also arXiv:1709.05845}\ }\BibitemShut {NoStop}%
\bibitem [{\citenamefont {Akrap}\ \emph {et~al.}(2016)\citenamefont {Akrap},
  \citenamefont {Hakl}, \citenamefont {Tchoumakov}, \citenamefont {Crassee},
  \citenamefont {Kuba}, \citenamefont {Goerbig}, \citenamefont {Homes},
  \citenamefont {Caha}, \citenamefont {Nov\'ak}, \citenamefont {Teppe},
  \citenamefont {Desrat}, \citenamefont {Koohpayeh}, \citenamefont {Wu},
  \citenamefont {Armitage}, \citenamefont {Nateprov}, \citenamefont
  {Arushanov}, \citenamefont {Gibson}, \citenamefont {Cava}, \citenamefont
  {van~der Marel}, \citenamefont {Piot}, \citenamefont {Faugeras},
  \citenamefont {Martinez}, \citenamefont {Potemski},\ and\ \citenamefont
  {Orlita}}]{w33}%
  \BibitemOpen
\bibfield  {journal} {  }\bibfield  {author} {\bibinfo {author} {\bibfnamefont
  {A.}~\bibnamefont {Akrap}}, \bibinfo {author} {\bibfnamefont
  {M.}~\bibnamefont {Hakl}}, \bibinfo {author} {\bibfnamefont {S.}~\bibnamefont
  {Tchoumakov}}, \bibinfo {author} {\bibfnamefont {I.}~\bibnamefont {Crassee}},
  \bibinfo {author} {\bibfnamefont {J.}~\bibnamefont {Kuba}}, \bibinfo {author}
  {\bibfnamefont {M.~O.}\ \bibnamefont {Goerbig}}, \bibinfo {author}
  {\bibfnamefont {C.~C.}\ \bibnamefont {Homes}}, \bibinfo {author}
  {\bibfnamefont {O.}~\bibnamefont {Caha}}, \bibinfo {author} {\bibfnamefont
  {J.}~\bibnamefont {Nov\'ak}}, \bibinfo {author} {\bibfnamefont
  {F.}~\bibnamefont {Teppe}}, \bibinfo {author} {\bibfnamefont
  {W.}~\bibnamefont {Desrat}}, \bibinfo {author} {\bibfnamefont
  {S.}~\bibnamefont {Koohpayeh}}, \bibinfo {author} {\bibfnamefont
  {L.}~\bibnamefont {Wu}}, \bibinfo {author} {\bibfnamefont {N.~P.}\
  \bibnamefont {Armitage}}, \bibinfo {author} {\bibfnamefont {A.}~\bibnamefont
  {Nateprov}}, \bibinfo {author} {\bibfnamefont {E.}~\bibnamefont {Arushanov}},
  \bibinfo {author} {\bibfnamefont {Q.~D.}\ \bibnamefont {Gibson}}, \bibinfo
  {author} {\bibfnamefont {R.~J.}\ \bibnamefont {Cava}}, \bibinfo {author}
  {\bibfnamefont {D.}~\bibnamefont {van~der Marel}}, \bibinfo {author}
  {\bibfnamefont {B.~A.}\ \bibnamefont {Piot}}, \bibinfo {author}
  {\bibfnamefont {C.}~\bibnamefont {Faugeras}}, \bibinfo {author}
  {\bibfnamefont {G.}~\bibnamefont {Martinez}}, \bibinfo {author}
  {\bibfnamefont {M.}~\bibnamefont {Potemski}}, \ and\ \bibinfo {author}
  {\bibfnamefont {M.}~\bibnamefont {Orlita}},\ }\href {\doibase
  10.1103/PhysRevLett.117.136401} {\bibfield  {journal} {\bibinfo  {journal}
  {Phys. Rev. Lett.}\ }\textbf {\bibinfo {volume} {117}},\ \bibinfo {pages}
  {136401} (\bibinfo {year} {2016})}\BibitemShut {NoStop}%
\bibitem [{\citenamefont {Desrat}\ \emph {et~al.}(2018)\citenamefont {Desrat},
  \citenamefont {Krishtopenko}, \citenamefont {Piot}, \citenamefont {Orlita},
  \citenamefont {Consejo}, \citenamefont {Ruffenach}, \citenamefont {Knap},
  \citenamefont {Nateprov}, \citenamefont {Arushanov},\ and\ \citenamefont
  {Teppe}}]{w34}%
  \BibitemOpen
  \bibfield  {author} {\bibinfo {author} {\bibfnamefont {W.}~\bibnamefont
  {Desrat}}, \bibinfo {author} {\bibfnamefont {S.~S.}\ \bibnamefont
  {Krishtopenko}}, \bibinfo {author} {\bibfnamefont {B.~A.}\ \bibnamefont
  {Piot}}, \bibinfo {author} {\bibfnamefont {M.}~\bibnamefont {Orlita}},
  \bibinfo {author} {\bibfnamefont {C.}~\bibnamefont {Consejo}}, \bibinfo
  {author} {\bibfnamefont {S.}~\bibnamefont {Ruffenach}}, \bibinfo {author}
  {\bibfnamefont {W.}~\bibnamefont {Knap}}, \bibinfo {author} {\bibfnamefont
  {A.}~\bibnamefont {Nateprov}}, \bibinfo {author} {\bibfnamefont
  {E.}~\bibnamefont {Arushanov}}, \ and\ \bibinfo {author} {\bibfnamefont
  {F.}~\bibnamefont {Teppe}},\ }\href {\doibase 10.1103/PhysRevB.97.245203}
  {\bibfield  {journal} {\bibinfo  {journal} {Phys. Rev. B}\ }\textbf {\bibinfo
  {volume} {97}},\ \bibinfo {pages} {245203} (\bibinfo {year}
  {2018})}\BibitemShut {NoStop}%
\bibitem [{\citenamefont {Malcolm}\ and\ \citenamefont {Nicol}(2015)}]{w28}%
  \BibitemOpen
  \bibfield  {author} {\bibinfo {author} {\bibfnamefont {J.~D.}\ \bibnamefont
  {Malcolm}}\ and\ \bibinfo {author} {\bibfnamefont {E.~J.}\ \bibnamefont
  {Nicol}},\ }\href {\doibase 10.1103/PhysRevB.92.035118} {\bibfield  {journal}
  {\bibinfo  {journal} {Phys. Rev. B}\ }\textbf {\bibinfo {volume} {92}},\
  \bibinfo {pages} {035118} (\bibinfo {year} {2015})}\BibitemShut {NoStop}%
\bibitem [{\citenamefont {Raoux}\ \emph {et~al.}(2014)\citenamefont {Raoux},
  \citenamefont {Morigi}, \citenamefont {Fuchs}, \citenamefont {Pi\'echon},\
  and\ \citenamefont {Montambaux}}]{w29}%
  \BibitemOpen
  \bibfield  {author} {\bibinfo {author} {\bibfnamefont {A.}~\bibnamefont
  {Raoux}}, \bibinfo {author} {\bibfnamefont {M.}~\bibnamefont {Morigi}},
  \bibinfo {author} {\bibfnamefont {J.-N.}\ \bibnamefont {Fuchs}}, \bibinfo
  {author} {\bibfnamefont {F.}~\bibnamefont {Pi\'echon}}, \ and\ \bibinfo
  {author} {\bibfnamefont {G.}~\bibnamefont {Montambaux}},\ }\href {\doibase
  10.1103/PhysRevLett.112.026402} {\bibfield  {journal} {\bibinfo  {journal}
  {Phys. Rev. Lett.}\ }\textbf {\bibinfo {volume} {112}},\ \bibinfo {pages}
  {026402} (\bibinfo {year} {2014})}\BibitemShut {NoStop}%
\bibitem [{\citenamefont {Wang}\ \emph {et~al.}(2013)\citenamefont {Wang},
  \citenamefont {Weng}, \citenamefont {Wu}, \citenamefont {Dai},\ and\
  \citenamefont {Fang}}]{w32}%
  \BibitemOpen
  \bibfield  {author} {\bibinfo {author} {\bibfnamefont {Z.}~\bibnamefont
  {Wang}}, \bibinfo {author} {\bibfnamefont {H.}~\bibnamefont {Weng}}, \bibinfo
  {author} {\bibfnamefont {Q.}~\bibnamefont {Wu}}, \bibinfo {author}
  {\bibfnamefont {X.}~\bibnamefont {Dai}}, \ and\ \bibinfo {author}
  {\bibfnamefont {Z.}~\bibnamefont {Fang}},\ }\href {\doibase
  10.1103/PhysRevB.88.125427} {\bibfield  {journal} {\bibinfo  {journal} {Phys.
  Rev. B}\ }\textbf {\bibinfo {volume} {88}},\ \bibinfo {pages} {125427}
  (\bibinfo {year} {2013})}\BibitemShut {NoStop}%
\bibitem [{\citenamefont {Chu}\ \emph {et~al.}(2011)\citenamefont {Chu},
  \citenamefont {Shan}, \citenamefont {Lu},\ and\ \citenamefont
  {Shen}}]{32HgTe}%
  \BibitemOpen
  \bibfield  {author} {\bibinfo {author} {\bibfnamefont {R.-L.}\ \bibnamefont
  {Chu}}, \bibinfo {author} {\bibfnamefont {W.-Y.}\ \bibnamefont {Shan}},
  \bibinfo {author} {\bibfnamefont {J.}~\bibnamefont {Lu}}, \ and\ \bibinfo
  {author} {\bibfnamefont {S.-Q.}\ \bibnamefont {Shen}},\ }\href {\doibase
  10.1103/PhysRevB.83.075110} {\bibfield  {journal} {\bibinfo  {journal} {Phys.
  Rev. B}\ }\textbf {\bibinfo {volume} {83}},\ \bibinfo {pages} {075110}
  (\bibinfo {year} {2011})}\BibitemShut {NoStop}%
\bibitem [{\citenamefont {Orlita}\ \emph {et~al.}(2014)\citenamefont {Orlita},
  \citenamefont {Basko}, \citenamefont {Zholudev}, \citenamefont {Teppe},
  \citenamefont {Knap}, \citenamefont {Gavrilenko}, \citenamefont {Mikhailov},
  \citenamefont {Dvoretskii}, \citenamefont {Neugebauer}, \citenamefont
  {Faugeras}, \citenamefont {Barra}, \citenamefont {Martinez},\ and\
  \citenamefont {Potemski}}]{w30}%
  \BibitemOpen
  \bibfield  {author} {\bibinfo {author} {\bibfnamefont {M.}~\bibnamefont
  {Orlita}}, \bibinfo {author} {\bibfnamefont {D.~M.}\ \bibnamefont {Basko}},
  \bibinfo {author} {\bibfnamefont {M.~S.}\ \bibnamefont {Zholudev}}, \bibinfo
  {author} {\bibfnamefont {F.}~\bibnamefont {Teppe}}, \bibinfo {author}
  {\bibfnamefont {W.}~\bibnamefont {Knap}}, \bibinfo {author} {\bibfnamefont
  {V.~I.}\ \bibnamefont {Gavrilenko}}, \bibinfo {author} {\bibfnamefont
  {N.~N.}\ \bibnamefont {Mikhailov}}, \bibinfo {author} {\bibfnamefont {S.~A.}\
  \bibnamefont {Dvoretskii}}, \bibinfo {author} {\bibfnamefont
  {P.}~\bibnamefont {Neugebauer}}, \bibinfo {author} {\bibfnamefont
  {C.}~\bibnamefont {Faugeras}}, \bibinfo {author} {\bibfnamefont {A.-L.}\
  \bibnamefont {Barra}}, \bibinfo {author} {\bibfnamefont {G.}~\bibnamefont
  {Martinez}}, \ and\ \bibinfo {author} {\bibfnamefont {M.}~\bibnamefont
  {Potemski}},\ }\href {\doibase 10.1038/nphys2857} {\bibfield  {journal}
  {\bibinfo  {journal} {Nature Phys.}\ }\textbf {\bibinfo {volume} {10}},\
  \bibinfo {pages} {233} (\bibinfo {year} {2014})}\BibitemShut {NoStop}%
\bibitem [{\citenamefont {Teppe}\ \emph {et~al.}(2016)\citenamefont {Teppe},
  \citenamefont {Marcinkiewicz}, \citenamefont {Krishtopenko}, \citenamefont
  {Ruffenach}, \citenamefont {Consejo}, \citenamefont {Kadykov}, \citenamefont
  {Desrat}, \citenamefont {But}, \citenamefont {Knap}, \citenamefont {Ludwig},
  \citenamefont {Moon}, \citenamefont {Smirnov}, \citenamefont {Orlita},
  \citenamefont {Jiang}, \citenamefont {Morozov}, \citenamefont {Gavrilenko},
  \citenamefont {Mikhailov},\ and\ \citenamefont {Dvoretskii}}]{w31}%
  \BibitemOpen
  \bibfield  {author} {\bibinfo {author} {\bibfnamefont {F.}~\bibnamefont
  {Teppe}}, \bibinfo {author} {\bibfnamefont {M.}~\bibnamefont
  {Marcinkiewicz}}, \bibinfo {author} {\bibfnamefont {S.~S.}\ \bibnamefont
  {Krishtopenko}}, \bibinfo {author} {\bibfnamefont {S.}~\bibnamefont
  {Ruffenach}}, \bibinfo {author} {\bibfnamefont {C.}~\bibnamefont {Consejo}},
  \bibinfo {author} {\bibfnamefont {A.~M.}\ \bibnamefont {Kadykov}}, \bibinfo
  {author} {\bibfnamefont {W.}~\bibnamefont {Desrat}}, \bibinfo {author}
  {\bibfnamefont {D.}~\bibnamefont {But}}, \bibinfo {author} {\bibfnamefont
  {W.}~\bibnamefont {Knap}}, \bibinfo {author} {\bibfnamefont {J.}~\bibnamefont
  {Ludwig}}, \bibinfo {author} {\bibfnamefont {S.}~\bibnamefont {Moon}},
  \bibinfo {author} {\bibfnamefont {D.}~\bibnamefont {Smirnov}}, \bibinfo
  {author} {\bibfnamefont {M.}~\bibnamefont {Orlita}}, \bibinfo {author}
  {\bibfnamefont {Z.}~\bibnamefont {Jiang}}, \bibinfo {author} {\bibfnamefont
  {S.~V.}\ \bibnamefont {Morozov}}, \bibinfo {author} {\bibfnamefont
  {V.}~\bibnamefont {Gavrilenko}}, \bibinfo {author} {\bibfnamefont {N.~N.}\
  \bibnamefont {Mikhailov}}, \ and\ \bibinfo {author} {\bibfnamefont {S.~A.}\
  \bibnamefont {Dvoretskii}},\ }\href {\doibase 10.1038/ncomms12576} {\bibfield
   {journal} {\bibinfo  {journal} {Nat. Commun.}\ }\textbf {\bibinfo {volume}
  {7}},\ \bibinfo {pages} {12576} (\bibinfo {year} {2016})}\BibitemShut
  {NoStop}%
\bibitem [{\citenamefont {Assaf}\ \emph {et~al.}(2016)\citenamefont {Assaf},
  \citenamefont {Phuphachong}, \citenamefont {Volobuev}, \citenamefont
  {Inhofer}, \citenamefont {Bauer}, \citenamefont {Springholz}, \citenamefont
  {de~Vaulchier},\ and\ \citenamefont {Guldner}}]{B1}%
  \BibitemOpen
  \bibfield  {author} {\bibinfo {author} {\bibfnamefont {B.}~\bibnamefont
  {Assaf}}, \bibinfo {author} {\bibfnamefont {T.}~\bibnamefont {Phuphachong}},
  \bibinfo {author} {\bibfnamefont {V.}~\bibnamefont {Volobuev}}, \bibinfo
  {author} {\bibfnamefont {A.}~\bibnamefont {Inhofer}}, \bibinfo {author}
  {\bibfnamefont {G.}~\bibnamefont {Bauer}}, \bibinfo {author} {\bibfnamefont
  {G.}~\bibnamefont {Springholz}}, \bibinfo {author} {\bibfnamefont
  {L.}~\bibnamefont {de~Vaulchier}}, \ and\ \bibinfo {author} {\bibfnamefont
  {Y.}~\bibnamefont {Guldner}},\ }\href {\doibase 10.1038/srep20323} {\bibfield
   {journal} {\bibinfo  {journal} {Sci. Rep.}\ }\textbf {\bibinfo {volume}
  {6}},\ \bibinfo {pages} {20323} (\bibinfo {year} {2016})}\BibitemShut
  {NoStop}%
\bibitem [{\citenamefont {Krizman}\ \emph {et~al.}(2018)\citenamefont
  {Krizman}, \citenamefont {Assaf}, \citenamefont {Phuphachong}, \citenamefont
  {Bauer}, \citenamefont {Springholz}, \citenamefont {de~Vaulchier},\ and\
  \citenamefont {Guldner}}]{B2}%
  \BibitemOpen
  \bibfield  {author} {\bibinfo {author} {\bibfnamefont {G.}~\bibnamefont
  {Krizman}}, \bibinfo {author} {\bibfnamefont {B.~A.}\ \bibnamefont {Assaf}},
  \bibinfo {author} {\bibfnamefont {T.}~\bibnamefont {Phuphachong}}, \bibinfo
  {author} {\bibfnamefont {G.}~\bibnamefont {Bauer}}, \bibinfo {author}
  {\bibfnamefont {G.}~\bibnamefont {Springholz}}, \bibinfo {author}
  {\bibfnamefont {L.~A.}\ \bibnamefont {de~Vaulchier}}, \ and\ \bibinfo
  {author} {\bibfnamefont {Y.}~\bibnamefont {Guldner}},\ }\href {\doibase
  10.1103/PhysRevB.98.245202} {\bibfield  {journal} {\bibinfo  {journal} {Phys.
  Rev. B}\ }\textbf {\bibinfo {volume} {98}},\ \bibinfo {pages} {245202}
  (\bibinfo {year} {2018})}\BibitemShut {NoStop}%
\bibitem [{\citenamefont {Ruan}\ \emph {et~al.}(2016)\citenamefont {Ruan},
  \citenamefont {Jian}, \citenamefont {Yao}, \citenamefont {Zhang},
  \citenamefont {Zhang},\ and\ \citenamefont {Xing}}]{w36a}%
  \BibitemOpen
  \bibfield  {author} {\bibinfo {author} {\bibfnamefont {J.}~\bibnamefont
  {Ruan}}, \bibinfo {author} {\bibfnamefont {S.-K.}\ \bibnamefont {Jian}},
  \bibinfo {author} {\bibfnamefont {H.}~\bibnamefont {Yao}}, \bibinfo {author}
  {\bibfnamefont {H.}~\bibnamefont {Zhang}}, \bibinfo {author} {\bibfnamefont
  {S.-C.}\ \bibnamefont {Zhang}}, \ and\ \bibinfo {author} {\bibfnamefont
  {D.}~\bibnamefont {Xing}},\ }\href {\doibase 10.1038/ncomms11136} {\bibfield
  {journal} {\bibinfo  {journal} {Nat. Commun.}\ }\textbf {\bibinfo {volume}
  {7}},\ \bibinfo {pages} {11136} (\bibinfo {year} {2016})}\BibitemShut
  {NoStop}%
\bibitem [{\citenamefont {Pfeffer}\ and\ \citenamefont
  {Zawadzki}(2003)}]{w2x2a}%
  \BibitemOpen
  \bibfield  {author} {\bibinfo {author} {\bibfnamefont {P.}~\bibnamefont
  {Pfeffer}}\ and\ \bibinfo {author} {\bibfnamefont {W.}~\bibnamefont
  {Zawadzki}},\ }\href {\doibase 10.1103/PhysRevB.68.035315} {\bibfield
  {journal} {\bibinfo  {journal} {Phys. Rev. B}\ }\textbf {\bibinfo {volume}
  {68}},\ \bibinfo {pages} {035315} (\bibinfo {year} {2003})}\BibitemShut
  {NoStop}%
\bibitem [{\citenamefont {Bernardes}\ \emph {et~al.}(2007)\citenamefont
  {Bernardes}, \citenamefont {Schliemann}, \citenamefont {Lee}, \citenamefont
  {Egues},\ and\ \citenamefont {Loss}}]{w2x2b}%
  \BibitemOpen
  \bibfield  {author} {\bibinfo {author} {\bibfnamefont {E.}~\bibnamefont
  {Bernardes}}, \bibinfo {author} {\bibfnamefont {J.}~\bibnamefont
  {Schliemann}}, \bibinfo {author} {\bibfnamefont {M.}~\bibnamefont {Lee}},
  \bibinfo {author} {\bibfnamefont {J.~C.}\ \bibnamefont {Egues}}, \ and\
  \bibinfo {author} {\bibfnamefont {D.}~\bibnamefont {Loss}},\ }\href {\doibase
  10.1103/PhysRevLett.99.076603} {\bibfield  {journal} {\bibinfo  {journal}
  {Phys. Rev. Lett.}\ }\textbf {\bibinfo {volume} {99}},\ \bibinfo {pages}
  {076603} (\bibinfo {year} {2007})}\BibitemShut {NoStop}%
\bibitem [{\citenamefont {Gavrilenko}\ \emph {et~al.}(2011)\citenamefont
  {Gavrilenko}, \citenamefont {Krishtopenko},\ and\ \citenamefont
  {Goiran}}]{w2x2c}%
  \BibitemOpen
  \bibfield  {author} {\bibinfo {author} {\bibfnamefont {V.~I.}\ \bibnamefont
  {Gavrilenko}}, \bibinfo {author} {\bibfnamefont {S.~S.}\ \bibnamefont
  {Krishtopenko}}, \ and\ \bibinfo {author} {\bibfnamefont {M.}~\bibnamefont
  {Goiran}},\ }\href {\doibase 10.1134/S1063782611010088} {\bibfield  {journal}
  {\bibinfo  {journal} {Semiconductors}\ }\textbf {\bibinfo {volume} {45}},\
  \bibinfo {pages} {110} (\bibinfo {year} {2011})}\BibitemShut {NoStop}%
\bibitem [{\citenamefont {Krishtopenko}\ \emph {et~al.}()\citenamefont
  {Krishtopenko}, \citenamefont {Gavrilenko},\ and\ \citenamefont
  {Goiran}}]{w2x2d}%
  \BibitemOpen
  \bibfield  {author} {\bibinfo {author} {\bibfnamefont {S.~S.}\ \bibnamefont
  {Krishtopenko}}, \bibinfo {author} {\bibfnamefont {V.~I.}\ \bibnamefont
  {Gavrilenko}}, \ and\ \bibinfo {author} {\bibfnamefont {M.}~\bibnamefont
  {Goiran}},\ }\href {\doibase 10.4028/www.scientific.net/SSP.190.554}
  {\bibfield  {journal} {\bibinfo  {journal} {Solid State Phenomena}\ }\textbf
  {\bibinfo {volume} {190}},\ \bibinfo {pages} {554}}\BibitemShut {NoStop}%
\bibitem [{\citenamefont {Bender}\ and\ \citenamefont {Boettcher}(1998)}]{NH1}%
  \BibitemOpen
  \bibfield  {author} {\bibinfo {author} {\bibfnamefont {C.~M.}\ \bibnamefont
  {Bender}}\ and\ \bibinfo {author} {\bibfnamefont {S.}~\bibnamefont
  {Boettcher}},\ }\href {\doibase 10.1103/PhysRevLett.80.5243} {\bibfield
  {journal} {\bibinfo  {journal} {Phys. Rev. Lett.}\ }\textbf {\bibinfo
  {volume} {80}},\ \bibinfo {pages} {5243} (\bibinfo {year}
  {1998})}\BibitemShut {NoStop}%
\bibitem [{\citenamefont {Bender}\ \emph {et~al.}(2002)\citenamefont {Bender},
  \citenamefont {Brody},\ and\ \citenamefont {Jones}}]{NH2}%
  \BibitemOpen
  \bibfield  {author} {\bibinfo {author} {\bibfnamefont {C.~M.}\ \bibnamefont
  {Bender}}, \bibinfo {author} {\bibfnamefont {D.~C.}\ \bibnamefont {Brody}}, \
  and\ \bibinfo {author} {\bibfnamefont {H.~F.}\ \bibnamefont {Jones}},\ }\href
  {\doibase 10.1103/PhysRevLett.89.270401} {\bibfield  {journal} {\bibinfo
  {journal} {Phys. Rev. Lett.}\ }\textbf {\bibinfo {volume} {89}},\ \bibinfo
  {pages} {270401} (\bibinfo {year} {2002})}\BibitemShut {NoStop}%
\bibitem [{\citenamefont {Bender}\ \emph {et~al.}(2007)\citenamefont {Bender},
  \citenamefont {Brody}, \citenamefont {Jones},\ and\ \citenamefont
  {Meister}}]{NH3}%
  \BibitemOpen
  \bibfield  {author} {\bibinfo {author} {\bibfnamefont {C.~M.}\ \bibnamefont
  {Bender}}, \bibinfo {author} {\bibfnamefont {D.~C.}\ \bibnamefont {Brody}},
  \bibinfo {author} {\bibfnamefont {H.~F.}\ \bibnamefont {Jones}}, \ and\
  \bibinfo {author} {\bibfnamefont {B.~K.}\ \bibnamefont {Meister}},\ }\href
  {\doibase 10.1103/PhysRevLett.98.040403} {\bibfield  {journal} {\bibinfo
  {journal} {Phys. Rev. Lett.}\ }\textbf {\bibinfo {volume} {98}},\ \bibinfo
  {pages} {040403} (\bibinfo {year} {2007})}\BibitemShut {NoStop}%
\bibitem [{\citenamefont {Konotop}\ \emph {et~al.}(2016)\citenamefont
  {Konotop}, \citenamefont {Yang},\ and\ \citenamefont {Zezyulin}}]{NH4}%
  \BibitemOpen
  \bibfield  {author} {\bibinfo {author} {\bibfnamefont {V.~V.}\ \bibnamefont
  {Konotop}}, \bibinfo {author} {\bibfnamefont {J.}~\bibnamefont {Yang}}, \
  and\ \bibinfo {author} {\bibfnamefont {D.~A.}\ \bibnamefont {Zezyulin}},\
  }\href {\doibase 10.1103/RevModPhys.88.035002} {\bibfield  {journal}
  {\bibinfo  {journal} {Rev. Mod. Phys.}\ }\textbf {\bibinfo {volume} {88}},\
  \bibinfo {pages} {035002} (\bibinfo {year} {2016})}\BibitemShut {NoStop}%
\bibitem [{\citenamefont {El-Ganainy}\ \emph {et~al.}(2018)\citenamefont
  {El-Ganainy}, \citenamefont {Makris}, \citenamefont {Khajavikhan},
  \citenamefont {Musslimani}, \citenamefont {Rotter},\ and\ \citenamefont
  {Christodoulides}}]{NH5}%
  \BibitemOpen
  \bibfield  {author} {\bibinfo {author} {\bibfnamefont {R.}~\bibnamefont
  {El-Ganainy}}, \bibinfo {author} {\bibfnamefont {K.~G.}\ \bibnamefont
  {Makris}}, \bibinfo {author} {\bibfnamefont {M.}~\bibnamefont {Khajavikhan}},
  \bibinfo {author} {\bibfnamefont {Z.~H.}\ \bibnamefont {Musslimani}},
  \bibinfo {author} {\bibfnamefont {S.}~\bibnamefont {Rotter}}, \ and\ \bibinfo
  {author} {\bibfnamefont {D.~N.}\ \bibnamefont {Christodoulides}},\ }\href
  {\doibase 10.1038/nphys4323} {\bibfield  {journal} {\bibinfo  {journal}
  {Nature Phys.}\ }\textbf {\bibinfo {volume} {14}},\ \bibinfo {pages} {11}
  (\bibinfo {year} {2018})}\BibitemShut {NoStop}%
\bibitem [{\citenamefont {Tchoumakov}\ \emph {et~al.}(2017)\citenamefont
  {Tchoumakov}, \citenamefont {Jouffrey}, \citenamefont {Inhofer},
  \citenamefont {Bocquillon}, \citenamefont {Pla\ifmmode~\mbox{\c{c}}\else
  \c{c}\fi{}ais}, \citenamefont {Carpentier},\ and\ \citenamefont
  {Goerbig}}]{VPStates}%
  \BibitemOpen
  \bibfield  {author} {\bibinfo {author} {\bibfnamefont {S.}~\bibnamefont
  {Tchoumakov}}, \bibinfo {author} {\bibfnamefont {V.}~\bibnamefont
  {Jouffrey}}, \bibinfo {author} {\bibfnamefont {A.}~\bibnamefont {Inhofer}},
  \bibinfo {author} {\bibfnamefont {E.}~\bibnamefont {Bocquillon}}, \bibinfo
  {author} {\bibfnamefont {B.}~\bibnamefont {Pla\ifmmode~\mbox{\c{c}}\else
  \c{c}\fi{}ais}}, \bibinfo {author} {\bibfnamefont {D.}~\bibnamefont
  {Carpentier}}, \ and\ \bibinfo {author} {\bibfnamefont {M.~O.}\ \bibnamefont
  {Goerbig}},\ }\href {\doibase 10.1103/PhysRevB.96.201302} {\bibfield
  {journal} {\bibinfo  {journal} {Phys. Rev. B}\ }\textbf {\bibinfo {volume}
  {96}},\ \bibinfo {pages} {201302} (\bibinfo {year} {2017})}\BibitemShut
  {NoStop}%
\bibitem [{\citenamefont {Shan}\ \emph {et~al.}(2010)\citenamefont {Shan},
  \citenamefont {Lu},\ and\ \citenamefont {Shen}}]{w41}%
  \BibitemOpen
  \bibfield  {author} {\bibinfo {author} {\bibfnamefont {W.-Y.}\ \bibnamefont
  {Shan}}, \bibinfo {author} {\bibfnamefont {H.-Z.}\ \bibnamefont {Lu}}, \ and\
  \bibinfo {author} {\bibfnamefont {S.-Q.}\ \bibnamefont {Shen}},\ }\href
  {\doibase 10.1088/1367-2630/12/4/043048} {\bibfield  {journal} {\bibinfo
  {journal} {New J. Phys.}\ }\textbf {\bibinfo {volume} {12}},\ \bibinfo
  {pages} {043048} (\bibinfo {year} {2010})}\BibitemShut {NoStop}%
\bibitem [{\citenamefont {Luttinger}(1956)}]{wDK2}%
  \BibitemOpen
  \bibfield  {author} {\bibinfo {author} {\bibfnamefont {J.~M.}\ \bibnamefont
  {Luttinger}},\ }\href {\doibase 10.1103/PhysRev.102.1030} {\bibfield
  {journal} {\bibinfo  {journal} {Phys. Rev.}\ }\textbf {\bibinfo {volume}
  {102}},\ \bibinfo {pages} {1030} (\bibinfo {year} {1956})}\BibitemShut
  {NoStop}%
\bibitem [{\citenamefont {Lin-Liu}\ and\ \citenamefont {Sham}(1985)}]{wDK3}%
  \BibitemOpen
  \bibfield  {author} {\bibinfo {author} {\bibfnamefont {Y.~R.}\ \bibnamefont
  {Lin-Liu}}\ and\ \bibinfo {author} {\bibfnamefont {L.~J.}\ \bibnamefont
  {Sham}},\ }\href {\doibase 10.1103/PhysRevB.32.5561} {\bibfield  {journal}
  {\bibinfo  {journal} {Phys. Rev. B}\ }\textbf {\bibinfo {volume} {32}},\
  \bibinfo {pages} {5561} (\bibinfo {year} {1985})}\BibitemShut {NoStop}%
\bibitem [{\citenamefont {Chang}\ \emph {et~al.}(1985)\citenamefont {Chang},
  \citenamefont {Schulman}, \citenamefont {Bastard}, \citenamefont {Guldner},\
  and\ \citenamefont {Voos}}]{wDK4}%
  \BibitemOpen
  \bibfield  {author} {\bibinfo {author} {\bibfnamefont {Y.-C.}\ \bibnamefont
  {Chang}}, \bibinfo {author} {\bibfnamefont {J.~N.}\ \bibnamefont {Schulman}},
  \bibinfo {author} {\bibfnamefont {G.}~\bibnamefont {Bastard}}, \bibinfo
  {author} {\bibfnamefont {Y.}~\bibnamefont {Guldner}}, \ and\ \bibinfo
  {author} {\bibfnamefont {M.}~\bibnamefont {Voos}},\ }\href {\doibase
  10.1103/PhysRevB.31.2557} {\bibfield  {journal} {\bibinfo  {journal} {Phys.
  Rev. B}\ }\textbf {\bibinfo {volume} {31}},\ \bibinfo {pages} {2557}
  (\bibinfo {year} {1985})}\BibitemShut {NoStop}%
\bibitem [{\citenamefont {Kozlov}\ \emph {et~al.}(2016)\citenamefont {Kozlov},
  \citenamefont {Bauer}, \citenamefont {Ziegler}, \citenamefont {Fischer},
  \citenamefont {Savchenko}, \citenamefont {Kvon}, \citenamefont {Mikhailov},
  \citenamefont {Dvoretsky},\ and\ \citenamefont {Weiss}}]{DF1}%
  \BibitemOpen
  \bibfield  {author} {\bibinfo {author} {\bibfnamefont {D.~A.}\ \bibnamefont
  {Kozlov}}, \bibinfo {author} {\bibfnamefont {D.}~\bibnamefont {Bauer}},
  \bibinfo {author} {\bibfnamefont {J.}~\bibnamefont {Ziegler}}, \bibinfo
  {author} {\bibfnamefont {R.}~\bibnamefont {Fischer}}, \bibinfo {author}
  {\bibfnamefont {M.~L.}\ \bibnamefont {Savchenko}}, \bibinfo {author}
  {\bibfnamefont {Z.~D.}\ \bibnamefont {Kvon}}, \bibinfo {author}
  {\bibfnamefont {N.~N.}\ \bibnamefont {Mikhailov}}, \bibinfo {author}
  {\bibfnamefont {S.~A.}\ \bibnamefont {Dvoretsky}}, \ and\ \bibinfo {author}
  {\bibfnamefont {D.}~\bibnamefont {Weiss}},\ }\href {\doibase
  10.1103/PhysRevLett.116.166802} {\bibfield  {journal} {\bibinfo  {journal}
  {Phys. Rev. Lett.}\ }\textbf {\bibinfo {volume} {116}},\ \bibinfo {pages}
  {166802} (\bibinfo {year} {2016})}\BibitemShut {NoStop}%
\bibitem [{\citenamefont {Thomas}\ \emph {et~al.}(2017)\citenamefont {Thomas},
  \citenamefont {Crauste}, \citenamefont {Haas}, \citenamefont {Jouneau},
  \citenamefont {B\"auerle}, \citenamefont {L\'evy}, \citenamefont {Orignac},
  \citenamefont {Carpentier}, \citenamefont {Ballet},\ and\ \citenamefont
  {Meunier}}]{DF2}%
  \BibitemOpen
  \bibfield  {author} {\bibinfo {author} {\bibfnamefont {C.}~\bibnamefont
  {Thomas}}, \bibinfo {author} {\bibfnamefont {O.}~\bibnamefont {Crauste}},
  \bibinfo {author} {\bibfnamefont {B.}~\bibnamefont {Haas}}, \bibinfo {author}
  {\bibfnamefont {P.-H.}\ \bibnamefont {Jouneau}}, \bibinfo {author}
  {\bibfnamefont {C.}~\bibnamefont {B\"auerle}}, \bibinfo {author}
  {\bibfnamefont {L.~P.}\ \bibnamefont {L\'evy}}, \bibinfo {author}
  {\bibfnamefont {E.}~\bibnamefont {Orignac}}, \bibinfo {author} {\bibfnamefont
  {D.}~\bibnamefont {Carpentier}}, \bibinfo {author} {\bibfnamefont
  {P.}~\bibnamefont {Ballet}}, \ and\ \bibinfo {author} {\bibfnamefont
  {T.}~\bibnamefont {Meunier}},\ }\href {\doibase 10.1103/PhysRevB.96.245420}
  {\bibfield  {journal} {\bibinfo  {journal} {Phys. Rev. B}\ }\textbf {\bibinfo
  {volume} {96}},\ \bibinfo {pages} {245420} (\bibinfo {year}
  {2017})}\BibitemShut {NoStop}%
\bibitem [{\citenamefont {Noel}\ \emph {et~al.}(2018)\citenamefont {Noel},
  \citenamefont {Thomas}, \citenamefont {Fu}, \citenamefont {Vila},
  \citenamefont {Haas}, \citenamefont {Jouneau}, \citenamefont {Gambarelli},
  \citenamefont {Meunier}, \citenamefont {Ballet},\ and\ \citenamefont
  {Attan\'e}}]{DF3}%
  \BibitemOpen
  \bibfield  {author} {\bibinfo {author} {\bibfnamefont {P.}~\bibnamefont
  {Noel}}, \bibinfo {author} {\bibfnamefont {C.}~\bibnamefont {Thomas}},
  \bibinfo {author} {\bibfnamefont {Y.}~\bibnamefont {Fu}}, \bibinfo {author}
  {\bibfnamefont {L.}~\bibnamefont {Vila}}, \bibinfo {author} {\bibfnamefont
  {B.}~\bibnamefont {Haas}}, \bibinfo {author} {\bibfnamefont {P.-H.}\
  \bibnamefont {Jouneau}}, \bibinfo {author} {\bibfnamefont {S.}~\bibnamefont
  {Gambarelli}}, \bibinfo {author} {\bibfnamefont {T.}~\bibnamefont {Meunier}},
  \bibinfo {author} {\bibfnamefont {P.}~\bibnamefont {Ballet}}, \ and\ \bibinfo
  {author} {\bibfnamefont {J.~P.}\ \bibnamefont {Attan\'e}},\ }\href {\doibase
  10.1103/PhysRevLett.120.167201} {\bibfield  {journal} {\bibinfo  {journal}
  {Phys. Rev. Lett.}\ }\textbf {\bibinfo {volume} {120}},\ \bibinfo {pages}
  {167201} (\bibinfo {year} {2018})}\BibitemShut {NoStop}%
\bibitem [{\citenamefont {Potter}\ \emph {et~al.}(2014)\citenamefont {Potter},
  \citenamefont {Kimchi},\ and\ \citenamefont {Vishwanath}}]{w36}%
  \BibitemOpen
  \bibfield  {author} {\bibinfo {author} {\bibfnamefont {A.~C.}\ \bibnamefont
  {Potter}}, \bibinfo {author} {\bibfnamefont {I.}~\bibnamefont {Kimchi}}, \
  and\ \bibinfo {author} {\bibfnamefont {A.}~\bibnamefont {Vishwanath}},\
  }\href {\doibase 10.1038/ncomms6161} {\bibfield  {journal} {\bibinfo
  {journal} {Nat. Commun.}\ }\textbf {\bibinfo {volume} {5}},\ \bibinfo {pages}
  {5161} (\bibinfo {year} {2014})}\BibitemShut {NoStop}%
\bibitem [{\citenamefont {Krishtopenko}\ \emph {et~al.}(2016)\citenamefont
  {Krishtopenko}, \citenamefont {Yahniuk}, \citenamefont {But}, \citenamefont
  {Gavrilenko}, \citenamefont {Knap},\ and\ \citenamefont {Teppe}}]{Kane}%
  \BibitemOpen
  \bibfield  {author} {\bibinfo {author} {\bibfnamefont {S.~S.}\ \bibnamefont
  {Krishtopenko}}, \bibinfo {author} {\bibfnamefont {I.}~\bibnamefont
  {Yahniuk}}, \bibinfo {author} {\bibfnamefont {D.~B.}\ \bibnamefont {But}},
  \bibinfo {author} {\bibfnamefont {V.~I.}\ \bibnamefont {Gavrilenko}},
  \bibinfo {author} {\bibfnamefont {W.}~\bibnamefont {Knap}}, \ and\ \bibinfo
  {author} {\bibfnamefont {F.}~\bibnamefont {Teppe}},\ }\href {\doibase
  10.1103/PhysRevB.94.245402} {\bibfield  {journal} {\bibinfo  {journal} {Phys.
  Rev. B}\ }\textbf {\bibinfo {volume} {94}},\ \bibinfo {pages} {245402}
  (\bibinfo {year} {2016})}\BibitemShut {NoStop}%
\end{thebibliography}
%

\end{document}